\documentclass[english,prd, 11pt, nofootinbib,  superscriptaddress, preprintnumbers,floatfix]{revtex4-2}
\usepackage[T1]{fontenc}
\usepackage[latin9]{inputenc}
\usepackage{babel}
\usepackage{mathrsfs}
\usepackage{bm,amsmath}
\usepackage{amssymb}

\usepackage{graphicx}
\usepackage{hyperref}
\hypersetup{pdftex,colorlinks=true,linkcolor=blue,citecolor=blue,menucolor=black,urlcolor=blue,filecolor=blue}

\providecommand{\tabularnewline}{\\}

\usepackage{slashed}
\usepackage{bbold}
\usepackage{multirow}

\usepackage{amsfonts}
\usepackage{makecell}
\usepackage[dvipsnames]{xcolor}
\usepackage[normalem]{ulem}
\usepackage{longtable}
\usepackage{array}
\usepackage{float}
\usepackage{subfigure}

\newcommand{\eq}{\begin{eqnarray}}
\newcommand{\en}{\end{eqnarray}}

\begin{document}
\title{Three-body coupled channel framework for two-neutron halo nuclei}
\author{Jin-Yi Pang}\email{jypang@usst.edu.cn}
\address{College of Science, University of Shanghai for Science and Technology,
Shanghai 200093, China}
\author{Li-Tan Li}\email{litanli480@gmail.com}
\address{College of Science, University of Shanghai for Science and Technology,
Shanghai 200093, China}
\author{Feng-Kun Guo}\email{Corresponding author. fkguo@itp.ac.cn}
\affiliation{CAS Key Laboratory of Theoretical Physics, Institute of Theoretical Physics, Chinese Academy of Sciences, Beijing 100190, China}
\affiliation{School of Physical Sciences, University of Chinese Academy of Sciences, Beijing 100049, China}
\affiliation{Peng Huanwu Collaborative Center for Research and Education, Beihang University, Beijing 100191, China}
\author{Jia-Jun Wu}\email{Corresponding author. wujiajun@ucas.ac.cn}
\address{School of Physical Sciences, University of Chinese Academy of Sciences,
Beijing 100049, China}

\begin{abstract}
We study the Borromean nuclei formed by a core nucleus and two neutrons in a  nonrelativistic effective field theory formalism considering both neutron-neutron and neutron-core interactions. 
We provide {formulae}  of the charge and matter radii, and successfully reproduce the universal relation proposed by Hongo and Son based on the approximation of an infinite neutron-neutron scattering length and neglecting the neutron-core scattering. 
Once the realistic finite neutron-neutron and neutron-core scattering lengths are used, the charge and matter radii are influenced by the neutron-core channel in a growingly relevant manner.  
We obtain a relation among the binding energy of the three-body Borromean system, the ratio between charge and matter radii, and the ratio between the neutron-neutron and core-neutron scattering lengths. We find that the two-neutron separation energy for $^{22}$C needs to be $\lesssim2$~keV in order to be consistent with the experimental constraints of the matter radius of  $^{22}$C and the $^{20}{\rm C}\,n$ $S$-wave scattering length.
\end{abstract}
\maketitle

\section{Introduction}

The Borromean nuclei have attracted more and more interests in recent years~\cite{Zhukov:1993aw,Hammer:2017tjm}.
It is a kind of halo nuclei including a relatively compact core and two neutrons that distribute around the core as a halo. 
The whole system can be interpreted as a three-body system.
Distinguished from normal three-body systems, the criterion of a Borromean system is that three particles form a bound state but any two-body subsystem cannot be bound. 
It has been known that $^{6}$He, $^{11}$Li and $^{22}$C all present such behavior in the nuclei spectrum, thus can be called Borromean nuclei.

Recently, in Ref.~\cite{Hongo:2022sdr}, the authors apply a nonrelativistic effective field theory (NREFT) to Borromean nuclei, and obtain a universal relation in the limit of infinite neutron-neutron scattering length ($a_{nn} \to \infty$) and zero core-neutron scattering length ($a_{An}\to 0$) as follows,
\begin{align}
\frac{\langle r_{m}^{2}\rangle}{\langle r_{c}^{2}\rangle} & =\frac{2}{3}A\,,\label{eq:annratioold}
\end{align}
where $\langle r_{m}^{2}\rangle$ and $\langle r_{c}^{2}\rangle$ are the matter and charge mean square radii, respectively, and $A$ is the nucleon number of the core nucleus inside the Borromean system. 
In such an NREFT, the power counting is based on that the absolute value of the $S$-wave neutron-neutron scattering length $a_{nn}$ is much larger than the core-neutron one $a_{An}$, and the neutron-core interaction would provide corrections in powers of $a_{An}/a_{nn}$. 

In this paper, we consider the effects of finite neutron-neutron
scattering length and corrections from the neutron-core interaction. 
We use the particle-dimer formalism developed for low-energy three-body interactions (for the dimer formalism, see Refs.~\cite{Kaplan:1996nv,Bedaque:1998kg,Bedaque:1998km}); such an NREFT is called halo effective field theory (halo EFT) for the study of halo nuclei~\cite{Bertulani:2002sz} (for its applications in three-body quantization conditions in a finite volume, see Refs.~\cite{Hammer:2017uqm,Hammer:2017kms,Doring:2018xxx,Pang:2019dfe,Pang:2020pkl,Muller:2021uur}). 
We consider not only the two-neutron dimer considered in Ref.~\cite{Hongo:2022sdr} but also the neutron-core dimer so that a complete description of the three-body Borromean nuclei is present dynamically in this work.
The two-body scattering can determine the parameters of involved dimer fields, while an additional three-body coupling is tuned to produce the binding energy of the shallow three-body bound state. 
Expressions for the matter and charge mean square radii of the Borromean nuclei will be derived. 
The ratio of the two mean-square radii is consistent with Eq.~\eqref{eq:annratioold} under the limit of $a_{nn} \gg a_{An} \sim 0$.
Within our formalism, corrections from finite values of $a_{nn}$ and $a_{An}$ can be specified, and are found to be sizeable in some cases. 
At last, the $^6$He, $^{11}$Li and $^{22}$C systems will be discussed.

The paper is organized as follows. 
Section~\ref{sec:Formalism} shows
the formalism we are using to study three-body systems. 
The form factors are calculated analytically in this section. 
A series of numerical computations
are present in Section~\ref{sec:Numerical-result}, 
and Section~\ref{sec:Conclusion}
gives our conclusion. 
Some details of the calculations are relegated to two appendices.

\section{\label{sec:Formalism}Formalism for three-body bound states}

\subsection{Lagrangian}

We use the particle-dimer formalism~\cite{Hammer:2017kms} to describe three-body bound systems of a core nucleus and two neutrons. 
Since two neutrons barely bind (there is a near-threshold virtual state pole) and the considered core nucleus and neutron do not bind either, the two-neutron halo nuclei in question are Borromean systems. 
To investigate the effects of the neutron-core ($An$) interactions and complete the physical picture of the three-body interactions, we consider both $nn$ and $An$ dimers, as auxiliary fields, though there is no two-body bound state. 
The effective Lagrangian is organized as follows, 
\begin{align}
\mathcal{L} & =\mathcal{L}_{1}+\mathcal{L}_{2}+\mathcal{L}_{3}\,,
\end{align}
where 
\begin{align}
\mathcal{L}_{1} & =n^{\dagger}\left(i\partial_{0}+\frac{\nabla^{2}}{2m}\right)n+A^{\dagger}\left(i\partial_{0}+\frac{\nabla^{2}}{2m_{A}}\right)A
+T_{nn}^{\dagger}\sigma_{nn}T_{nn}+T_{An}^{\dagger}\sigma_{An}T_{An},
 \\
\mathcal{L}_{2} & =\frac{1}{2}T_{nn}^{\dagger}nn+T_{An}^{\dagger}An+\text{H.c.}\,, \label{eq:L2}\\
\mathcal{L}_{3} & =h_{0}\big[T_{nn}^{\dagger}A^{\dagger}\big]\big[T_{nn}A\big]\,, \label{eq:L3}
\end{align}
with $m$ and $m_A$ the masses of the neutron and the core nucleus, respectively.
The kinematic part $\mathcal{L}_{1}$ includes single particle fields
$n$ and $A$ which denote the neutron and the core, respectively.
The $nn$ dimer and $An$ dimer are denoted as the auxiliary fields $T_{nn}$ and $T_{An}$, respectively. 
The two-body dynamics can be described by the interaction between the dimer fields and their constituents, i.e.,  the coupling of the $T_{nn}$ dimer to two neutrons $nn$ and the coupling of the $T_{An}$ dimer to the core $A$ and neutron $n$. 
Here, the two-body couplings have been absorbed into the parameters $\sigma_{nn}$ and $\sigma_{An}$ in $\mathcal{L}_1$, so we are allowed to set the couplings to unity in $\mathcal{L}_2$ at the leading order; the factor of $1/2$ in the first term of $\mathcal{L}_2$ is to account for identical neutrons.
The short-distance three-body interaction is described by $\mathcal{L}_{3}$. 
Formally, there could be three terms for the three-body interactions, corresponding to different combinations of the dimer and particle fields,
\begin{align}
h_{0} & \left(T_{nn}^{\dagger}A^{\dagger}\right)\left(T_{nn}A\right)+h_{0}^{'}\left(T_{An}^{\dagger}n^{\dagger}\right)\left(T_{An}n\right)
+h_{0}^{''}\left[\left(T_{nn}^{\dagger}A^{\dagger}\right)\left(T_{An}n\right)+\text{H.c.}\right].\label{three-operator}
\end{align}
However, these terms are not independent. 
We can integrate out the auxiliary dimer fields and reproduce the three-body interaction in the equivalent effective field theory without dimers, $h(A^{\dagger}n^{\dagger}n^{\dagger})(Ann)$, which has only a single term at the leading order. 
The relation between the general low energy constant (LEC), $h$, and the LECs, $h_{0},h_{0}^{'}$ and $h_{0}^{''}$
is, therefore, given by
\begin{align}
h & =\frac{h_{0}}{\sigma_{nn}^{2}}+\frac{h_{0}^{'}}{\sigma_{An}^{2}}+\frac{2h_{0}^{''}}{\sigma_{nn}\sigma_{An}}.
\end{align}
Thus we are allowed to choose $h_{0}^{'}=h_{0}^{''}=0$ and use the
contact term in $(nn)+A\to(nn)+A$ in Eq.~\eqref{three-operator} to parameterize the three-body
force. 
That is to say, any of the operators in Eq.~\eqref{three-operator} can describe the three-body dynamics at the leading order by considering that the dimer field is equivalent to the two particles coupled to it. 

\subsection{Scattering equation}

\begin{figure}
\begin{centering}
\includegraphics[scale=0.25]{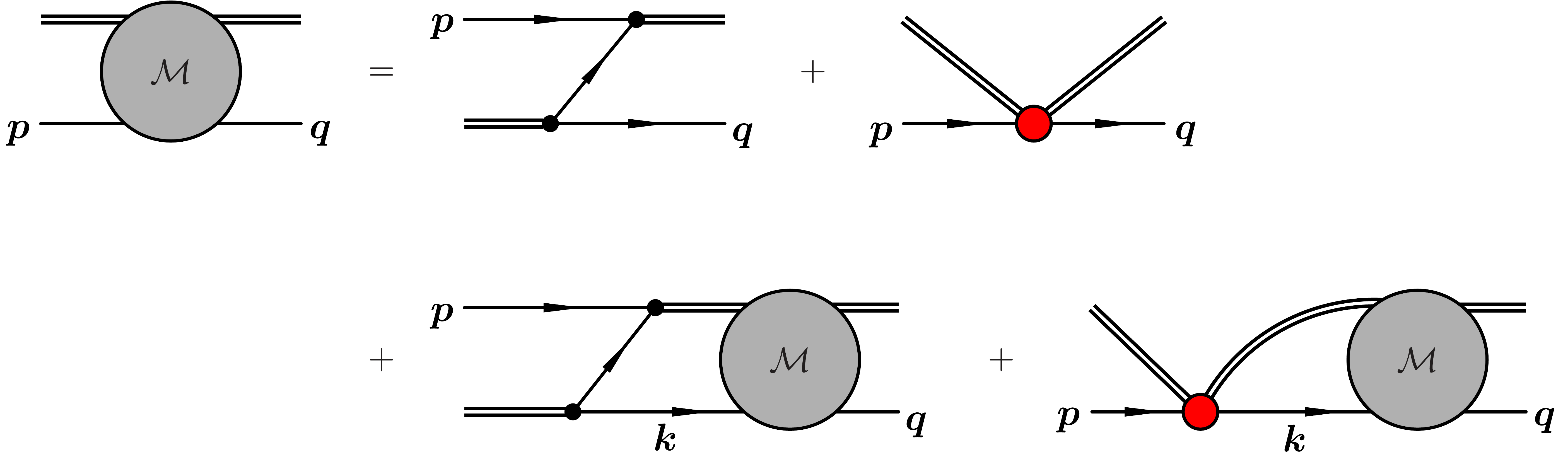}
\par\end{centering}
\caption{\label{fig:scattering-equation} Diagrammatic representation of the scattering equation in the particle-dimer formalism. 
The double lines denote the dimers  and the single lines are for the spectators. }
\end{figure}

The information of the $Ann$ Borromean system is encoded in scattering
amplitude $Ann\to Ann$. 
In the particle-dimer formalism~\cite{Zhukov:1993aw,Hammer:2017tjm}, there are two scattering channels, i.e., the $nn$ dimer plus the core $A$ (denoted as the $nn$ channel), and the $An$ dimer plus a spectator $n$ (denoted as the $An$
channel). 
Correspondingly, the Faddeev equation of the three-body scattering is as follows, 
\begin{align}
    \mathcal{M}(\bm{p},\bm{q};E) & =Z(\bm{p},\bm{q};E)+\int^{\Lambda}\frac{d^{3}\bm k}{(2\pi)^{3}}Z(\bm{p},\bm{k};E)\tau(k;E)\mathcal{M}(\bm{k},\bm{q};E), \label{eq:scattering}
\end{align}
where $\bm{p}$ ($\bm{q}, \bm{k}$) is the three-momentum of the spectator in the center-of-mass (c.m.) frame of the three-body system, $p$ ($q, k$) is its magnitude, $E$ is the total energy of the three-body system in the same frame defined relative to the three-body threshold, and $\Lambda$ is the hard cutoff used to regularize the integral scattering equation by truncating the momentum of the spectator. 
In the above equation, each term is  a $2\times2$ matrix in the channel space. 
The scattering equation is shown diagrammatically in Fig.~\ref{fig:scattering-equation}.

The scattering amplitude $\mathcal{M}$ takes the form of 
\begin{align}
\mathcal{M} & =\begin{pmatrix}\mathcal{M}_{11} & \mathcal{M}_{12}\\
\mathcal{M}_{21} & \mathcal{M}_{22}
\end{pmatrix}.
\end{align}
The subscript $1$ and $2$ represent $An$ and $nn$ channels, respectively.
The off-diagonal terms are thus the amplitudes of the pertinent cross
channels. 

The dimer propagator $\tau$ is in two channels as well, given by the diagonal matrix,
\begin{align}
\tau & =\begin{pmatrix}\tau_{1}\\
 & \tau_{2}
\end{pmatrix}.
\end{align}
In each channel, the dimer propagator matches the $S$-wave two-body scattering amplitude as
\begin{align}
\tau_{1}(p;E) & =\frac{2\pi}{\mu}\frac{1}{-a_{An}^{-1}+\sqrt{c_{An}^{2}p^{2}-2\mu E}}, \label{eq:tau1}\\
\tau_{2}(p;E) & =\frac{8\pi}{m}\frac{1}{-a_{nn}^{-1}+\sqrt{c_{nn}^{2}p^{2}-mE}},\label{eq:tau2}
\end{align}
where $\mu$ is the reduced mass of the $An$ subsystem, 
$E$ is the total energy of the three-body system, and $p$ is the magnitude of the three-momentum of the corresponding spectator in the c.m. frame of the three-body system.
Here, we are using the effective range expansion in the following form, 
\begin{align}
k_0\cot\delta & =-\frac{1}{a}+\frac{1}{2}r k_0^2+\cdots,
\end{align}
where $\delta$ is the two-body $S$-wave scattering phase shift, $k_0$ is the c.m. momentum, and $a$ and $r$ are the scattering length and effective range, respectively. 
So $a_{An(nn)}$ is two-body scattering length of $A$ and $n$ ($n$
and $n$). 
The two-body interaction parameters $\sigma_{nn}$ and $\sigma_{An}$ are matched to the corresponding scattering lengths $a_{nn}$ and $a_{An}$ using dimensional regularization with the minimal subtraction scheme as follows,
\begin{align}
\frac{2\pi\sigma_{An}}{\mu} & =-\dfrac{1}{a_{An}}\,,\quad \frac{8\pi\sigma_{nn}}{m} =-\frac{1}{a_{nn}}\,.
\end{align}
The coefficients $c_{An}$ and $c_{nn}$ in Eqs.~\eqref{eq:tau1} and \eqref{eq:tau2} are defined as
\begin{align}
c_{An}^{2} & =1-\frac{\mu^{2}}{m_{A}^{2}}\,,\quad c_{nn}^{2}=\frac{m}{2m_{A}}+\frac{1}{4}\,.
\end{align}
One can verify that the square roots in Eqs.~\eqref{eq:tau1} and \eqref{eq:tau2} give the relative momenta (differ by an imaginary $i$) between two particles inside the dimer in their respective c.m. frames.

The kernel of the scattering equation~\eqref{eq:scattering} $Z$ is a $2\times2$ matrix given by
\begin{align}
Z & =\begin{pmatrix}Z_{11} & Z_{12}\\
Z_{21} & Z_{22}
\end{pmatrix}.
\end{align}
It includes both the long-range interactions via the exchange neutron or core and the short-range three-body interaction. 
The introduction of the three-body contact term in Eq.~\eqref{eq:L3} implies that a non-zero contact term is present only in $Z_{22}$.
Therefore, we have 
\begin{align}
Z_{11}(\bm{p},\bm{q};E) & =\frac{2\mu}{p^{2}+q^{2}+\frac{2\mu}{m_{A}}\bm{p}\cdot\bm{q}-2\mu E}\,,\\
Z_{12}(\bm{p},\bm{q};E)=Z_{21}(\bm{q},\bm{p};E) & =\frac{m}{p^{2}+\frac{m}{2\mu}q^{2}+\bm{p}\cdot\bm{q}-mE}\,,\\
Z_{22}(\bm{p},\bm{q};E) & =h_0\equiv\frac{H_{0}}{\Lambda^{2}}\,.\label{eq:3-body-force}
\end{align}
Renormalization group invariance implies that the short-range coupling $H_{0}$ runs with the scale $\Lambda$, i.e., $H_{0}=H_{0}(\Lambda)$.
Equation~\eqref{eq:scattering} can be projected onto the $S$-wave, leading to the following isotropic kernels, 
\begin{align}
Z_{11}(p,q;E) & =\frac{m_{A}}{2pq}\log\left(\frac{p^{2}+q^{2}+\frac{2\mu}{m_{A}}pq-2\mu E}{p^{2}+q^{2}-\frac{2\mu}{m_{A}}pq-2\mu E}\right),\\
Z_{12}(p,q;E)=Z_{21}(q,p;E) & =\frac{m}{2pq}\log\left(\frac{p^{2}+\frac{m}{2\mu}q^{2}+pq-mE}{p^{2}+\frac{m}{2\mu}q^{2}-pq-mE}\right),\\
Z_{22}(p,q;E) & =\frac{H_{0}}{\Lambda^{2}}\,.
\end{align}

\subsection{Bound state solution of the Faddeev equation}

The $Ann$ three-body bound state corresponds to a pole below the three-body threshold of the particle-dimer scattering amplitude. 
Thus, we have
\begin{align}
\mathcal{M}(\bm{p},\bm{q};E) & =-\frac{\psi(\bm{p})\psi^{\dagger}(\bm{q})}{E+B}+\text{regular terms},\label{eq:particle-dimer-Amp}
\end{align}
where $B$ is the three-body binding energy, $\psi$ is the particle-dimer
wave function that is dimensionless. 
Based on the scattering equation, we can write down the homogeneous equation at the bound state pole, i.e., $E=-B$,
\begin{align}
\psi(\bm{p}) & =\int^{\Lambda}\frac{d^{3}k}{(2\pi)^{3}}Z(\bm{p},\bm{k};E)\tau(k;E)\psi(\bm{k})\,.
\end{align}
The corresponding normalization is (for detailed derivations, see Appendix~\ref{sec:Normalization-of-wave function}), 
\begin{align}
\int\frac{d^{3}\bm{p}}{(2\pi)^{3}}
\frac{d^{3}\bm{q}}{(2\pi)^{3}}
\psi^{\dagger}(\bm{p})
\left[\tau(p;E)
\left(\frac{\partial}{\partial E}
Z(p,q;E)\right)
\tau(q;E)\right]
\Bigg|_{E=-B}\psi(\bm q)
\nonumber \\
+\int\frac{d^{3}\bm{p}}{(2\pi)^{3}}
\psi^{\dagger}(\bm{p})
\left(\frac{\partial}{\partial E}
\Bigg|_{E=-B}
\tau(p;E)\right)\psi(\bm{p})=1\,.
\end{align}

Since we have considered both the $An$ and $nn$ channels, the wave function $\psi$
has two components in the channel space, $\psi=(\psi_{1},\psi_{2})^{\text{T}}$, where $\psi_{1(2)}$ describes the $An$ ($nn$) channel.
The two wave functions at $\bm{p}=0$ give the couplings of the three-body bound state to the two particle-dimer channels, that is
\begin{align}
G_{An}=\psi_{1}(0),\quad G_{nn} & =\psi_{2}(0). \label{eq:couplings}
\end{align}
Here we pick up the coupling at the leading order in the low momentum expansion. 

\subsection{Form factors and the corresponding radii}

In order to analyze the form factors of Borromean nuclei, we introduce a physical bound state field $\mathcal{T}$ and rewrite the three-body Lagrangian $\mathcal{L}_{3}$ in the following way,
\begin{align}
\mathcal{L}_{3} & =\mathcal{T}^{\dagger}\sigma_{3}\left(i\partial_{t}+\frac{\nabla^{2}}{2M}+B_{b}\right)\mathcal{T}+g_{nn}\left(\mathcal{T}^{\dagger}T_{nn}A+\text{H.c.}\right)+g_{An}\left(\mathcal{T}^{\dagger}T_{An}n+\text{H.c.}\right),\label{eq:borromean-Lagrangian}
\end{align}
where $B_b$ is the bare binding energy, and $\sigma_3$ is a three-body coupling parameter.
By matching the above Lagrangian 
to the particle-dimer scattering amplitude (\ref{eq:particle-dimer-Amp}), we find that the new couplings $g_{nn}$ and $g_{An}$ are exactly the residues $G_{nn}$ and $G_{An}$ at the leading order, respectively, i.e., 
\begin{align}
g_{nn}=G_{nn},\quad & g_{An}=G_{An}.
\end{align}
The parameters $\sigma_{3}$ and $B_{b}$ can be fixed by the three-body binding energy $B$ and the field strength normalization condition, i.e., the residue of trimer propagator is unity. 
Thus, we have 
\begin{align}
\sigma_{3} =&\,1-4\pi g_{nn}^{2}\int^{\Lambda}\frac{d^{3}q}{(2\pi)^{3}}\frac{1}{\sqrt{c_{nn}^{2}q^{2}+mB}}\frac{1}{(-a_{nn}^{-1}+\sqrt{c_{nn}^{2}q^{2}+mB})^{2}}\nonumber \\
 & -2\pi g_{An}^{2}\int^{\Lambda}\frac{d^{3}q}{(2\pi)^{3}}\frac{1}{\sqrt{c_{An}^{2}q^{2}+2\mu B}}\frac{1}{(-a_{An}^{-1}+\sqrt{c_{An}^{2}q^{2}+2\mu B})^{2}},\\
B_{b} =&\,B-\frac{1}{\sigma_{3}}\Bigg[\frac{8\pi g_{nn}^{2}}{m}\int^{\Lambda}\frac{d^{3}q}{(2\pi)^{3}}\frac{1}{-a_{nn}^{-1}+\sqrt{c_{nn}^{2}q^{2}+mB}}\nonumber \\
 & +\frac{2\pi g_{An}^{2}}{\mu}\int^{\Lambda}\frac{d^{3}q}{(2\pi)^{3}}\frac{1}{-a_{An}^{-1}+\sqrt{c_{An}^{2}q^{2}+2\mu B}}\Bigg].
\end{align}
Here we have used the momentum cutoff $\Lambda$ to regularize the ultraviolet divergence as before.

\begin{figure}[tbh]
\centering
\subfigure[]{
\label{fig:charge-form-factor}
\includegraphics[scale=0.3]{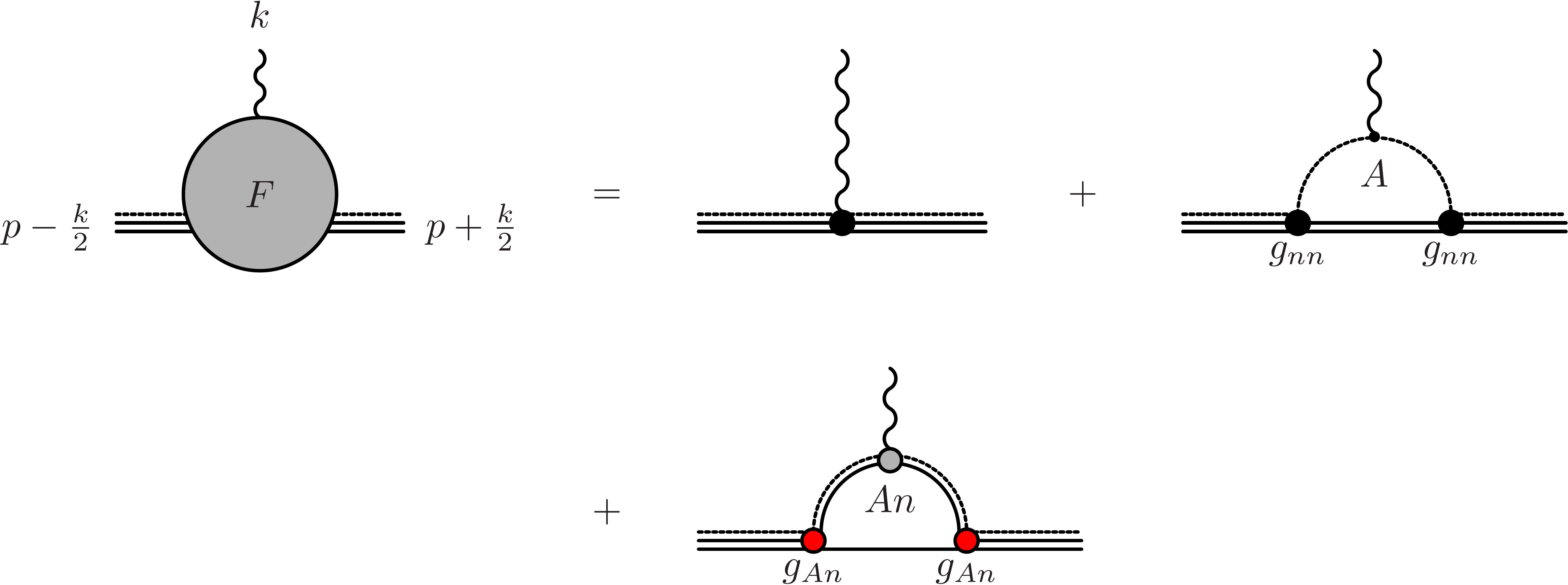}}
\\
\subfigure[]{
\label{fig:neutron-form-factor}
\includegraphics[scale=0.3]{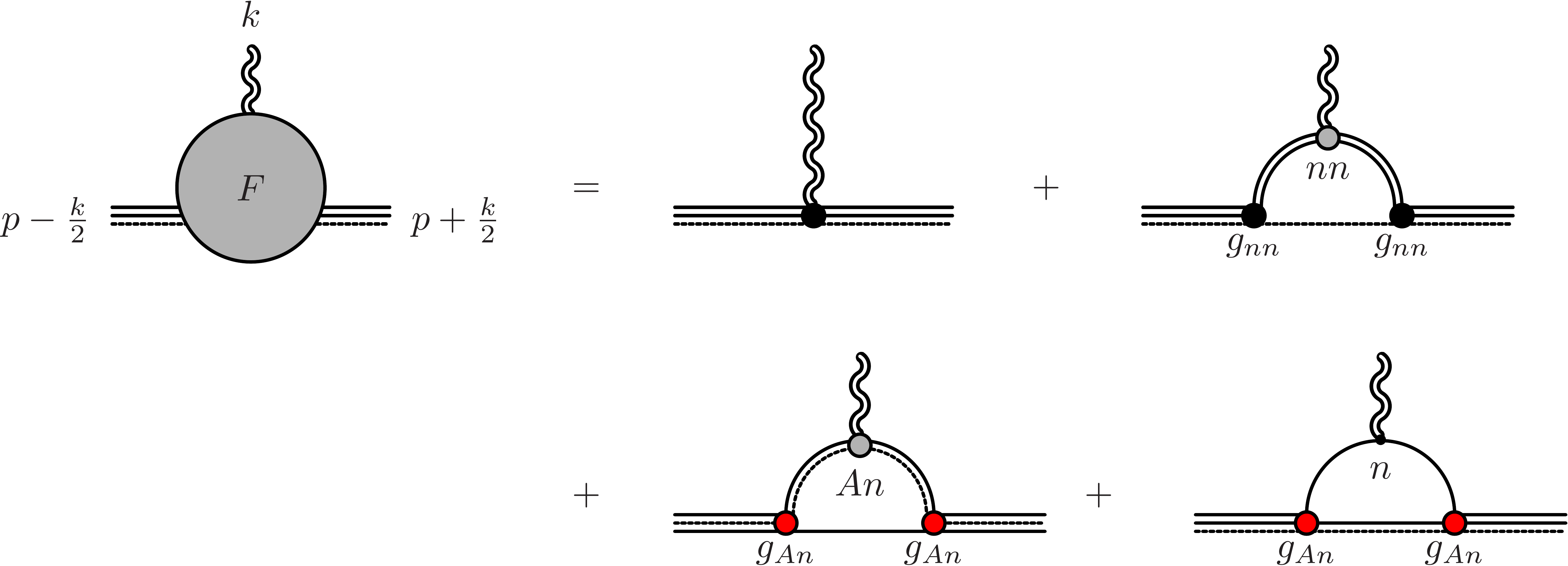}}
\caption{
(a) Charge form factor; 
(b) neutron form factor. 
Here we use wiggly lines to denote the electromagnetic current and doubly wiggly lines to denote the neutron number operator.}
\end{figure}

The charge form factor is extracted from the vertex function of bound
state $\mathcal{T}$ coupled to the electromagnetic current (see Fig.~\ref{fig:charge-form-factor}),
\begin{align}
F(k) & =1-\frac{1}{6}\langle r_{c}^{2}\rangle k^{2}+O(k^{4}),
\end{align}
where $k$ represents the four-momentum carried by the external current.
After evaluating the loop diagrams and performing the low momentum expansion, there are two contributions to the charge mean square radius (called radius for simplicity in the following) from the $nn$ and $An$ channels, respectively (some details of the derivations are given in Appendix~\ref{app:ff}), 
\begin{align}
\langle r_{c}^{2}\rangle & =\langle r_{c,nn}^{2}\rangle+\langle r_{c,An}^{2}\rangle,\\
\langle r_{c,nn}^{2}\rangle & =\frac{2G_{nn}^{2}}{\pi mB}\frac{4A^{1/2}}{(A+2)^{5/2}}  f_{c}(\gamma_{nn}) ,\label{eq:rcnn}\\
\langle r_{c,An}^{2}\rangle & =\frac{G_{An}^{2}}{\pi mB} \frac{4A^{1/2}}{(A+2)^{5/2}} \frac{(A+1)^{2}}{16A^{2}}\Bigg[\left(\frac{A+2}{A}\right)f_{n}(\gamma_{An})+f_{c}(\gamma_{An})\Bigg].\label{eq:rcAn}
\end{align}
Here we have approximated the mass of the core nucleus as $m_{A}=Am$ and the mass of the two-neutron halo nucleus as $M=(A+2)m$. 
The functions $f_n$ and $f_c$ are defined as
\begin{align}
f_{c}(\gamma) & =\frac{3}{4}I_{2,\frac{3}{2},\frac{1}{2}}+\frac{3}{2}I_{3,1,\frac{1}{2}}-\frac{1}{4}I_{2,\frac{5}{2},\frac{3}{2}}-\frac{1}{2}I_{3,2,\frac{3}{2}}-\frac{1}{2}I_{4,\frac{3}{2},\frac{3}{2}},\\
f_{n}(\gamma) & =\frac{1}{2}I_{2,\frac{3}{2},\frac{1}{2}},
\end{align}
where 
\begin{align}
I_{a,b,c}(\gamma) & =\int_{1}^{\infty}(y-1)^{c} \,\frac{t_{\gamma}^{a}(y)}{y^{b}} \,dy,\quad t_{\gamma}(y)=\frac{1}{\sqrt{y}+1/\gamma}.
\end{align}
The quantity $\gamma$ in Eqs.~(\ref{eq:rcnn}) and (\ref{eq:rcAn}) is defined from the two-body scattering length and the three-body binding energy as 
\begin{align}
\gamma_{nn}=-a_{nn}\sqrt{mB}, & \quad\gamma_{An}=-a_{An}\sqrt{2\mu B}.
\end{align}

In addition, we can also calculate the neutron radius via the coupling of the two-neutron nucleus to the neutron number operator (see Fig.~\ref{fig:neutron-form-factor}).
Analogously, the radius is obtained from the low-momentum expansion of the form factor,
\begin{align}
F_{n}(k) & =2\left[1-\frac{1}{6}\langle r_{n}^{2}\rangle k^{2}+O(k^{4})\right].
\end{align}
The factor $2$ represents the two neutrons in the system. 
There are two contributions as well,
\begin{align}
\langle r_{n}^{2}\rangle & =\langle r_{n,nn}^{2}\rangle+\langle r_{n,An}^{2}\rangle,\\
\langle r_{n,nn}^{2}\rangle & =\frac{2G_{nn}^{2}}{\pi mB}\left(\frac{A}{A+2}\right)^{3/2}\Bigg\{ f_{n}(\gamma_{nn})+\frac{A}{A+2}f_{c}(\gamma_{nn})\Bigg\},\label{eq:rnnn}\\
\langle r_{n,An}^{2}\rangle & =\frac{G_{An}^{2}}{\pi mB}\left(\frac{A}{A+2}\right)^{3/2}\frac{(A+1)^{2}}{8A^{2}}\Bigg\{ f_{n}(\gamma_{An})+\left[1+\frac{2}{A(A+2)}\right]f_{c}(\gamma_{An})\Bigg\}.\label{eq:rnAn}
\end{align}

In the above calculation, the charge radius reflects the distribution of the core nucleus, which is positively charged, and the neutron radius is for the distribution of the two neutrons in the system.
The matter radius of the two-neutron halo nucleus can be obtained as~\cite{Hammer:2017tjm,Hongo:2022sdr}
\begin{align}
\langle r_{m}^{2}\rangle & =\frac{2}{A+2}\langle r_{n}^{2}\rangle+\frac{A}{A+2}\langle r_{c}^{2}\rangle.
\end{align}
Consequently, we have
\begin{align}
\frac{\langle r_{m}^{2}\rangle}{A\langle r_{c}^{2}\rangle} & =\frac{1}{2}\left\{1+\frac{f_{n}(\gamma_{nn})+\frac{\kappa^{2}}{32}\left(\frac{A+1}{A}\right)^{2}\Big[\frac{A+2}{A}f_{c}(\gamma_{An})+f_{n}(\gamma_{An})\Big]}{f_{c}(\gamma_{nn})+\frac{\kappa^{2}}{32}\left(\frac{A+1}{A}\right)^{2}\Big[f_{c}(\gamma_{An})+\frac{A+2}{A}f_{n}(\gamma_{An})\Big]}\right\},
\end{align}
where $\kappa \equiv{G_{An}}/{G_{nn}}$.
It can be checked that in the unitary limit of the $nn$ channel, the universal relation in Ref.~\cite{Hongo:2022sdr} is achieved by taking $\kappa\to 0$, 
\begin{align}
\lim_{\gamma_{nn}\to\infty}\lim_{\kappa\to 0}\frac{\langle r_{m}^{2}\rangle}{A\langle r_{c}^{2}\rangle} & =\lim_{\gamma_{nn}\to\infty}\frac{1}{2}\Bigg[1+\frac{f_{n}(\gamma_{nn})}{f_{c}(\gamma_{nn})}\Bigg]=\frac{2}{3}.
\end{align}

\section{Numerical results}
\label{sec:Numerical-result}

\subsection{Quantitative analysis of a toy system}

\begin{figure}[tb]
\begin{centering}
\subfigure[]{\includegraphics[width=0.45\textwidth]{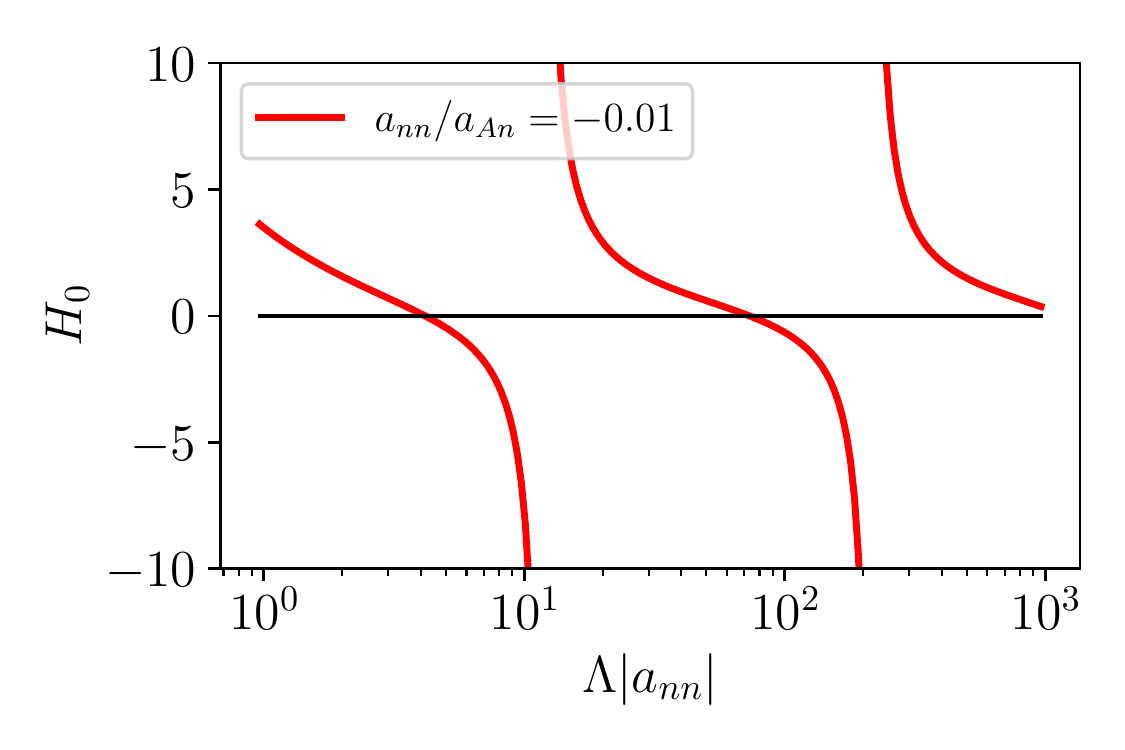} 

}$\qquad$\subfigure[]{\includegraphics[width=0.45\textwidth]{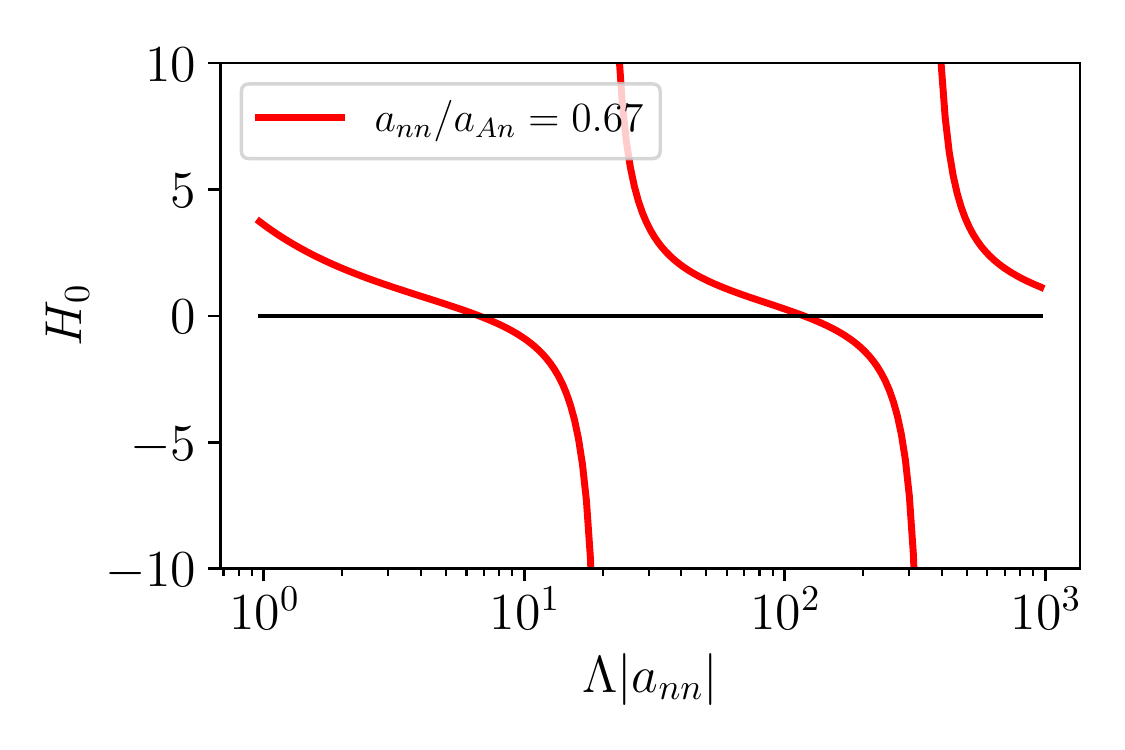} 

}
\par\end{centering}
\begin{centering}
\subfigure[]{\includegraphics[width=0.45\textwidth]{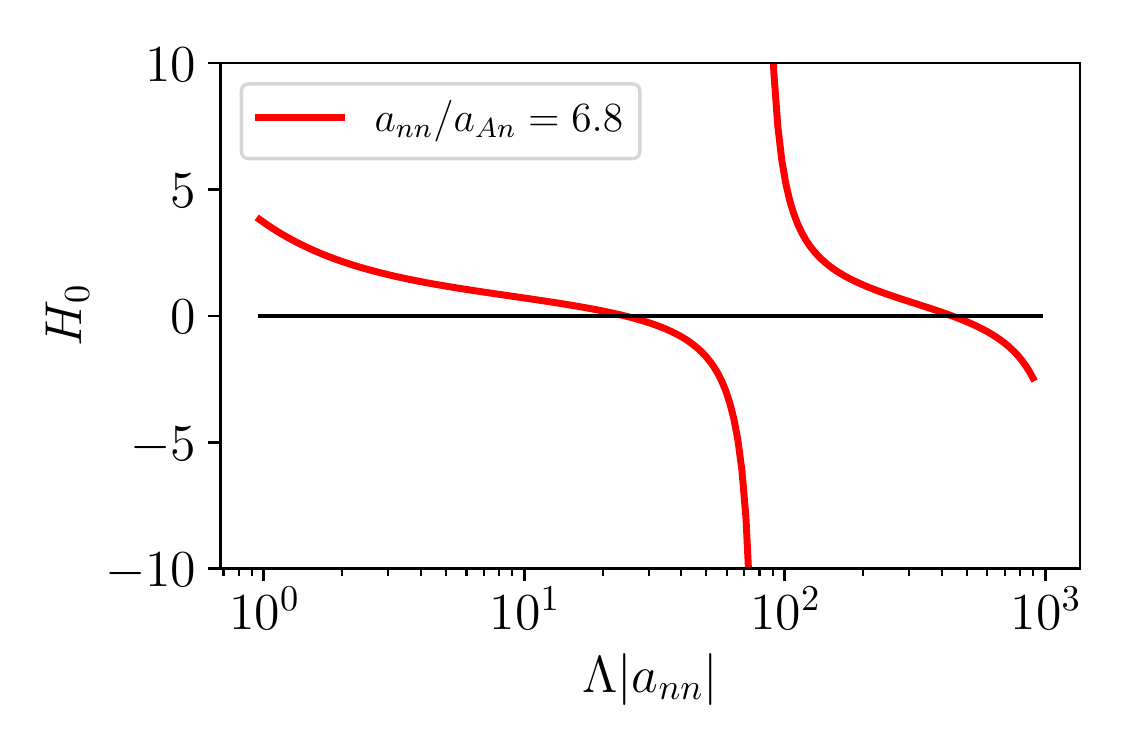} 

}$\qquad$\subfigure[]{\includegraphics[width=0.45\textwidth]{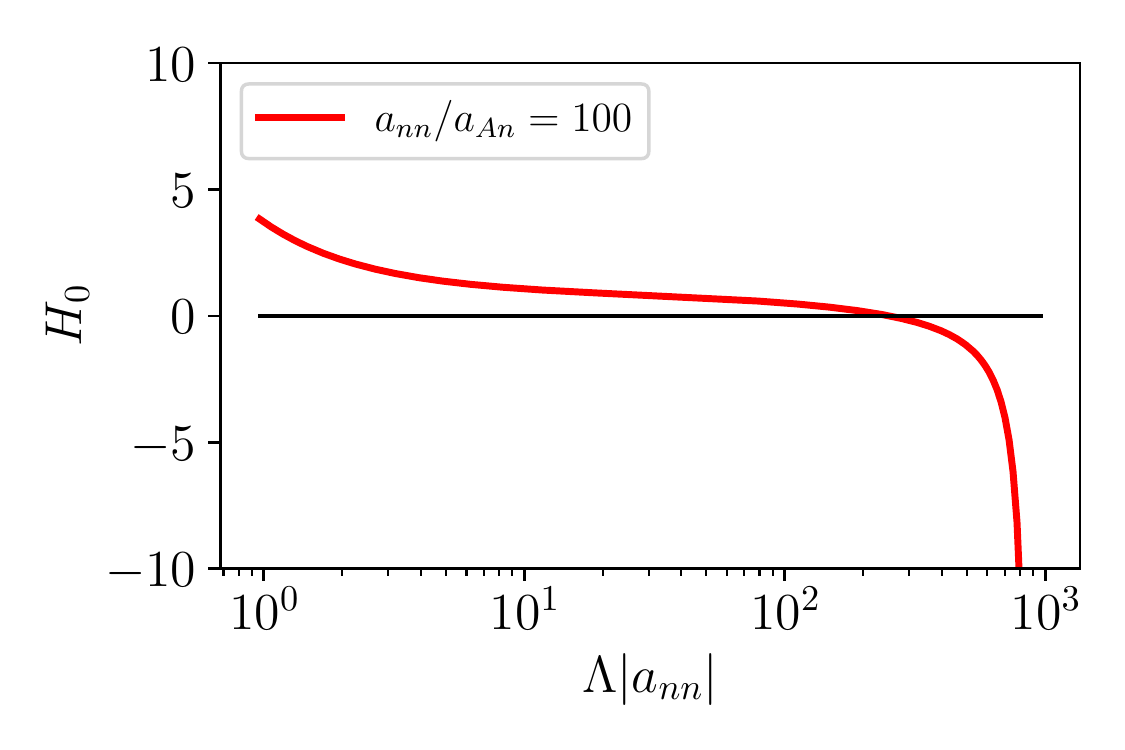} 

}
\par\end{centering}
\caption{\label{fig:3-body-force}The $\Lambda$ dependence of the short-range coupling with $a_{nn}=-18.6$~fm and different $a_{An}$ values.
(a) $a_{nn}/a_{An}=-0.01$; (b) $a_{nn}/a_{An}=0.67$; (c) $a_{nn}/a_{An}=6.8$;
(d) $a_{nn}/a_{An}=100$. They all support a three-body bound state
with binding energy $B=20$~keV. The mass number $A$ is set
to be $20$. }
\end{figure}

The $nn$ scattering in $^{1}S_{0}$ is known to be close to the unitary limit with a large (for the absolute value) scattering length $a_{nn}=-18.6$~fm~\cite{Chen:2008zzj}. 
We consider the scattering length
in the $An$ channel as a tuning parameter. 
A shallow three-body bound state can always be produced by varying the short-range three-body contact term in Eq.~(\ref{eq:3-body-force}).
The running coupling constant
shows a logarithmically periodic behavior similar to Efimov states; see Fig.~\ref{fig:3-body-force}, which is plotted for a two-neutron halo nucleus, a toy system with binding energy $B=20$~keV and mass number of the core $A=20$.
It can be expected that the system has Efimov states exactly when both
$a_{An}$ and $a_{nn}$ approach the unitary limit.  
For the case of Fig.~\ref{fig:3-body-force} (a), we set $a_{nn}/a_{An} = -0.01$, meaning that $a_{An}$ is positive, very close to the unitary limit, and the $An$ subsystem forms a two-body bound state; the three-body bound state is not a Borromean system. 
For the other plots in Fig.~\ref{fig:3-body-force}, the values of $a_{An}$ are negative, and the three-body bound states are of the Borromean type.

\begin{figure}[tb]
\begin{centering}
\subfigure[]{\includegraphics[width=0.45\textwidth]{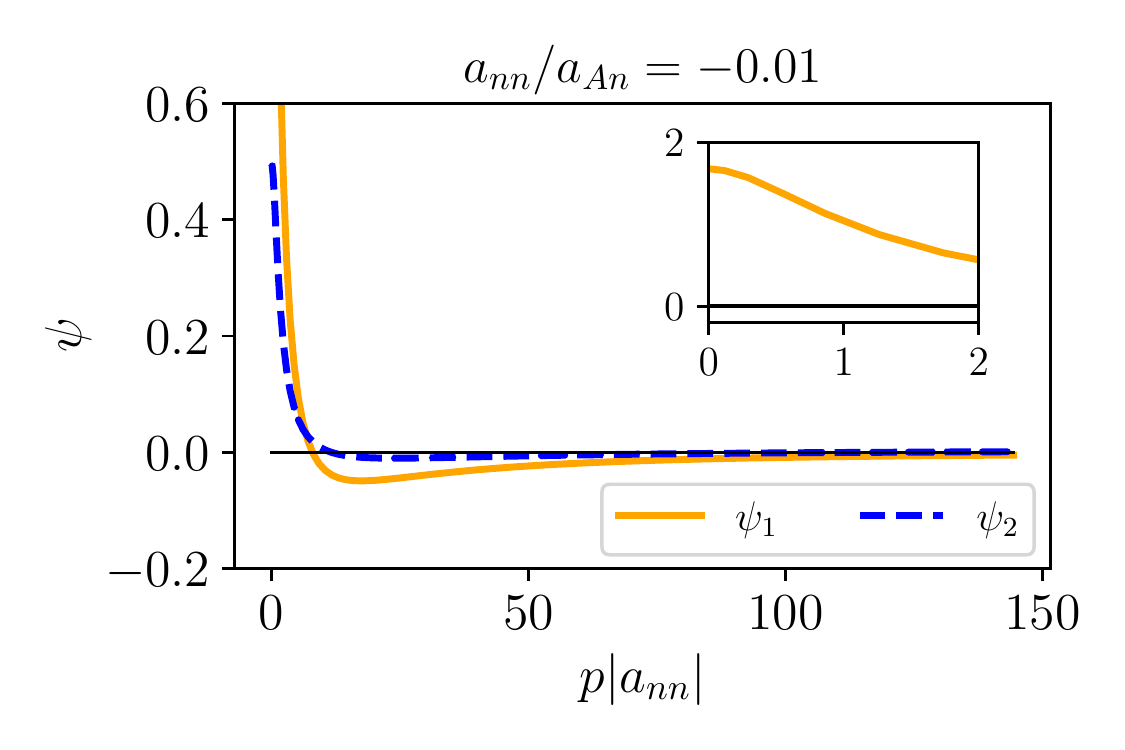} 

}$\qquad$\subfigure[]{\includegraphics[width=0.45\textwidth]{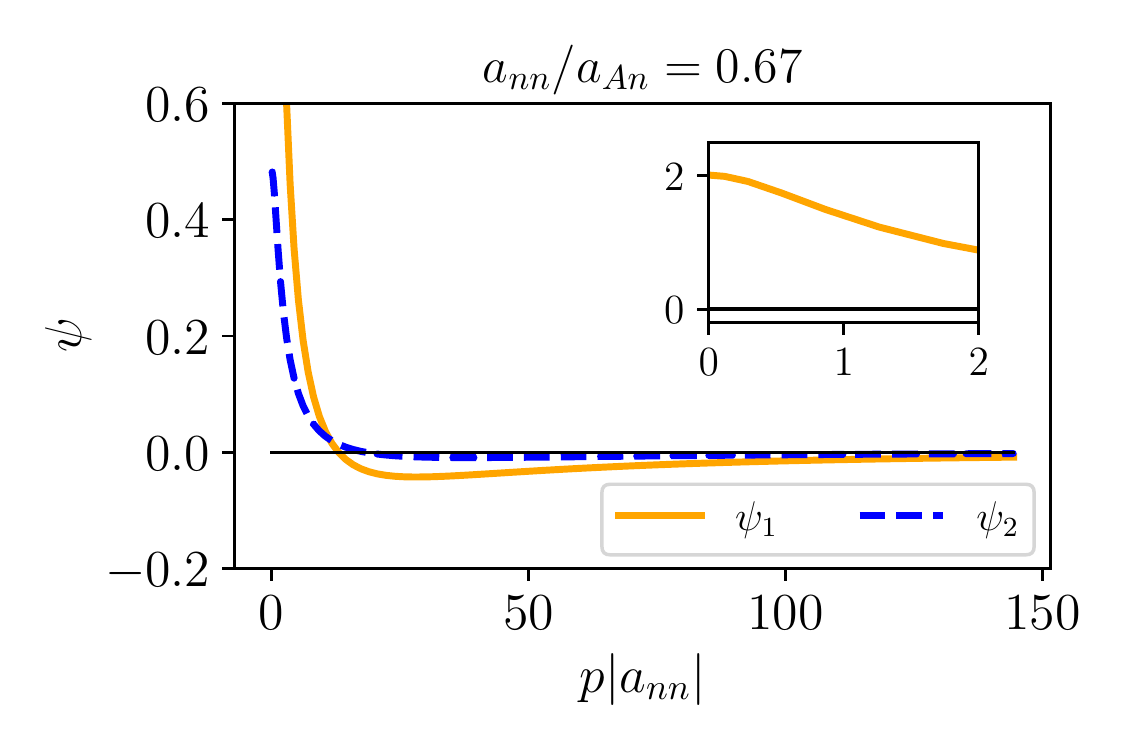} 

}
\par\end{centering}
\begin{centering}
\subfigure[]{\includegraphics[width=0.45\textwidth]{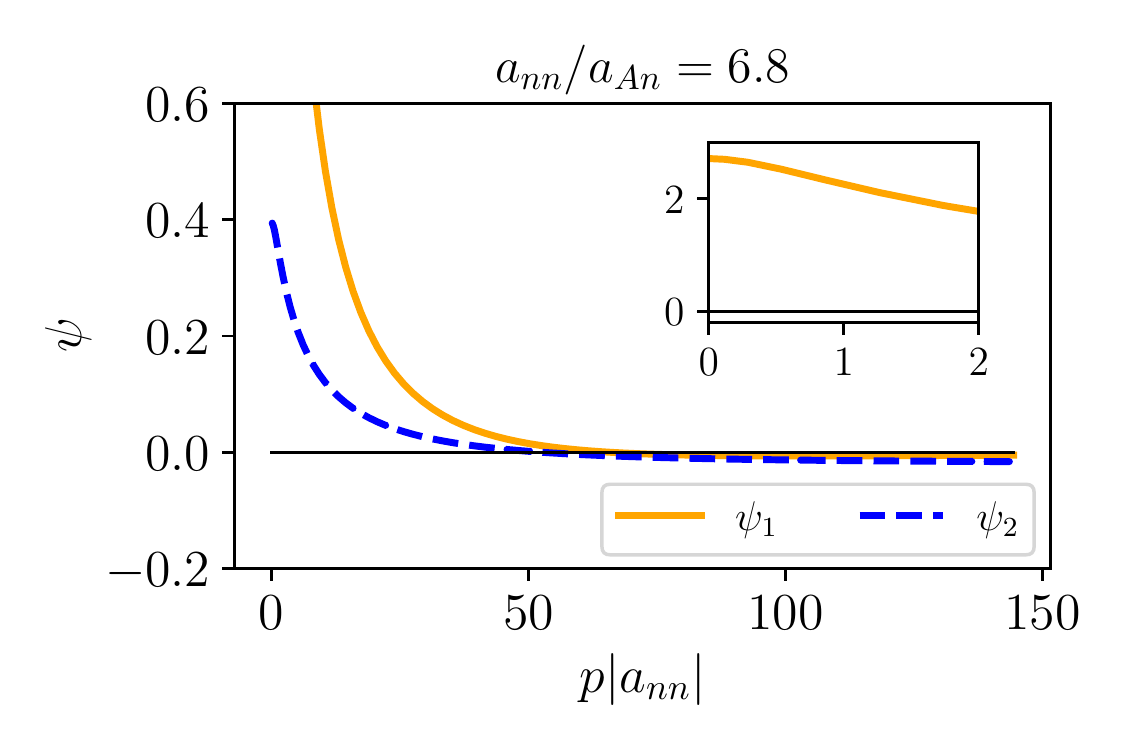} 

}$\qquad$\subfigure[]{\includegraphics[width=0.45\textwidth]{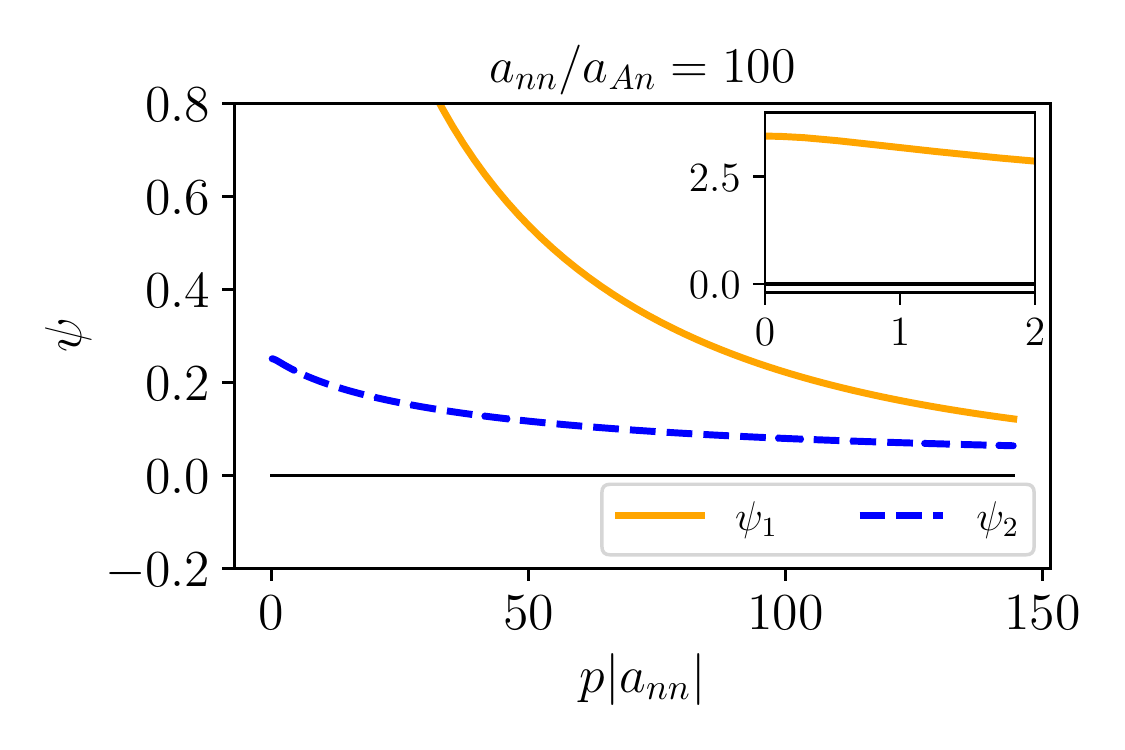} 

}
\par\end{centering}
\caption{\label{fig:wave function}Particle-dimer wave functions of the two-neutron halo nucleus for different $a_{An}$ values. (a) $a_{nn}/a_{An}=-0.01$; (b) $a_{nn}/a_{An}=0.67$;
(c) $a_{nn}/a_{An}=6.8$; (d) $a_{nn}/a_{An}=100$. Here we set $A=20$, $B=20$~keV and $a_{nn}=-18.6$~fm.}
\end{figure}

\begin{figure}
\begin{centering}
\includegraphics[scale=0.5]{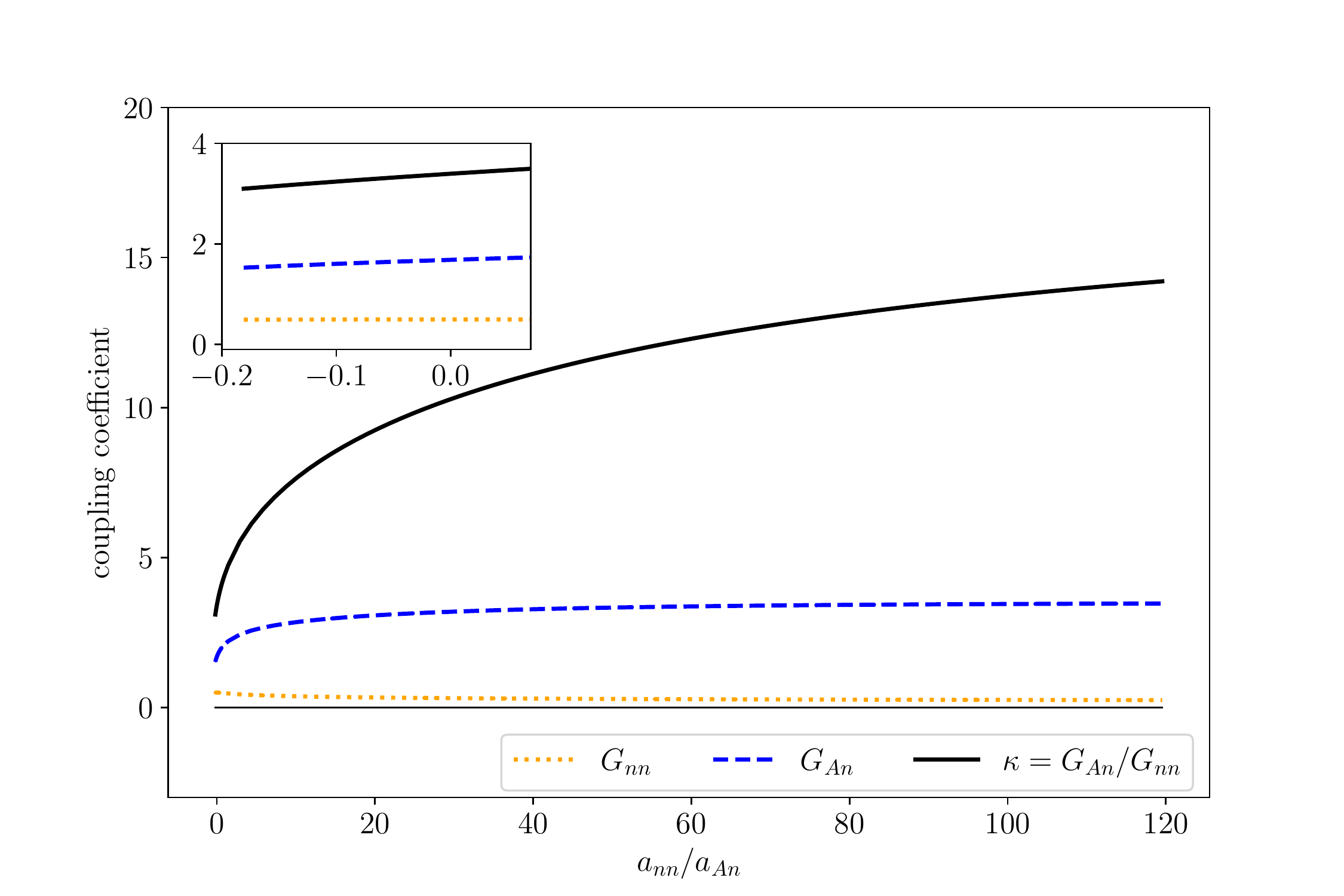}
\par\end{centering}
\caption{\label{fig:channel-coupling-ratio}Couplings of the two-neutron halo nucleus to the $An$ and $nn$ channels and their ratio. Here we set $A=20$, $B=20$~keV and $a_{nn}=-18.6$~fm.
}
\end{figure}

The particle-dimer wave functions in both the $nn$ and $An$ channels for a two-neutron halo nucleus with the same $B=20$~keV and $A=20$ as in Fig.~\ref{fig:3-body-force} are shown for different values of $a_{An}$ in Fig.~\ref{fig:wave function}, where $a_{nn}$ is taken to be $-18.6$~fm. 
It is interesting to notice that measuring by the channel couplings defined in Eq.~\eqref{eq:couplings}, which are the wave functions at vanishing momenta, the two-halo nuclei in question always couple more strongly to the $An$ channel than to the $nn$ one. 
The couplings $G_{An}$ and $G_{nn}$ and their ratio $\kappa=G_{A n} / G_{n n}$ for the case of $B=20$~keV, $A=20$ and $a_{nn}=-18.6$~fm with different values of $a_{An}$ are shown in  Table~\ref{tab:channel-coupling} and Fig.~\ref{fig:channel-coupling-ratio}. 
We check that the same pattern also holds if using the binding energies and mass numbers of the $^6$He and $^{11}$Li.

\begin{table}[tb] 
\begin{centering}
\caption{\label{tab:channel-coupling}Couplings of the two-neutron halo nucleus
to the $An$ and $nn$ channels and their ratios. Here we set $A=20$, $B=20$~keV and $a_{nn}=-18.6$~fm.
}
\begin{ruledtabular}
\begin{tabular}{c|cccc}
$a_{nn}/a_{An}$ & $-0.01$ & $0.67$ & $6.8$ & $100$\tabularnewline
\hline 
\hline 
$\gamma_{nn}=a_{nn}\sqrt{mB}$ & \multicolumn{4}{c}{$0.417$}\tabularnewline
$\gamma_{An}=a_{An}\sqrt{2\mu B}$ & $-17.9$ & $0.267$ & $0.0263$ & $0.00179$\tabularnewline
\hline 
$G_{nn}$ & $0.496$ & $0.484$ & $0.395$ & $0.251$\tabularnewline
$G_{An}$ & $1.68$ & $2.01$ & $2.71$ & $3.45$\tabularnewline
$\kappa=G_{An}/G_{nn}$ & $3.38$ & $4.15$ & $6.87$ & $13.7$\tabularnewline
\end{tabular}
\end{ruledtabular}
\par\end{centering}
\end{table}

\begin{table}[tb]
    \begin{centering}
    \caption{\label{tab:radius}Charge and matter rms radii of the two-neutron halo nucleus. Here we set $A=20$, $B=20$~keV and $a_{nn}=-18.6$~fm. 
    The subscripts denote the corresponding channels.
    }
        \begin{ruledtabular}
    \begin{tabular}{c|cccc}
    $a_{nn}/a_{An}$ & $-0.01$ & $0.67$ & $6.8$ & $100$\tabularnewline
    \hline 
    \hline 
    $\sqrt{\langle r_{c}^{2}\rangle_{nn}}$ (fm) & $0.602$ & $0.588$ & $0.479$ & $0.305$\tabularnewline
    $\sqrt{\langle r_{c}^{2}\rangle}=\sqrt{\langle r_{c}^{2}\rangle_{nn}+\langle r_{c}^{2}\rangle_{An}}$
    (fm) & $1.291$ & $0.974$ & $0.551$ & $0.307$\tabularnewline
    \hline 
    $\sqrt{\langle r_{n}^{2}\rangle_{nn}}$ (fm) & $7.473$ & $7.300$ & $5.915$ & $3.788$\tabularnewline
    $\sqrt{\langle r_{n}^{2}\rangle}=\sqrt{\langle r_{n}^{2}\rangle_{nn}+\langle r_{n}^{2}\rangle_{An}}$
    (fm) & $18.960$ & $13.705$ & $7.095$ & $3.822$\tabularnewline
    \hline 
    $\sqrt{\langle r_{m}^{2}\rangle_{nn}}$ (fm) & $2.325$ & $2.271$ & $1.852$ & $1.178$\tabularnewline
    $\sqrt{\langle r_{m}^{2}\rangle}=\sqrt{\langle r_{m}^{2}\rangle_{nn}+\langle r_{m}^{2}\rangle_{An}}$
    (fm) & $5.848$ & $4.235$ & $2.203$ & $1.189$\tabularnewline
    \hline 
    $\langle r_{m}^{2}\rangle_{nn}\big/(A\langle r_{c}^{2}\rangle_{nn})$  & \multicolumn{4}{c}{$0.746$}\tabularnewline
    $\langle r_{m}^{2}\rangle\big/(A\langle r_{c}^{2}\rangle)$  & $1.025$ & $0.945$ & $0.798$ & $0.748$\tabularnewline
    \end{tabular}
\end{ruledtabular}
    \par\end{centering}
    \end{table}

\begin{figure}
\begin{centering}
\includegraphics[scale=0.5]{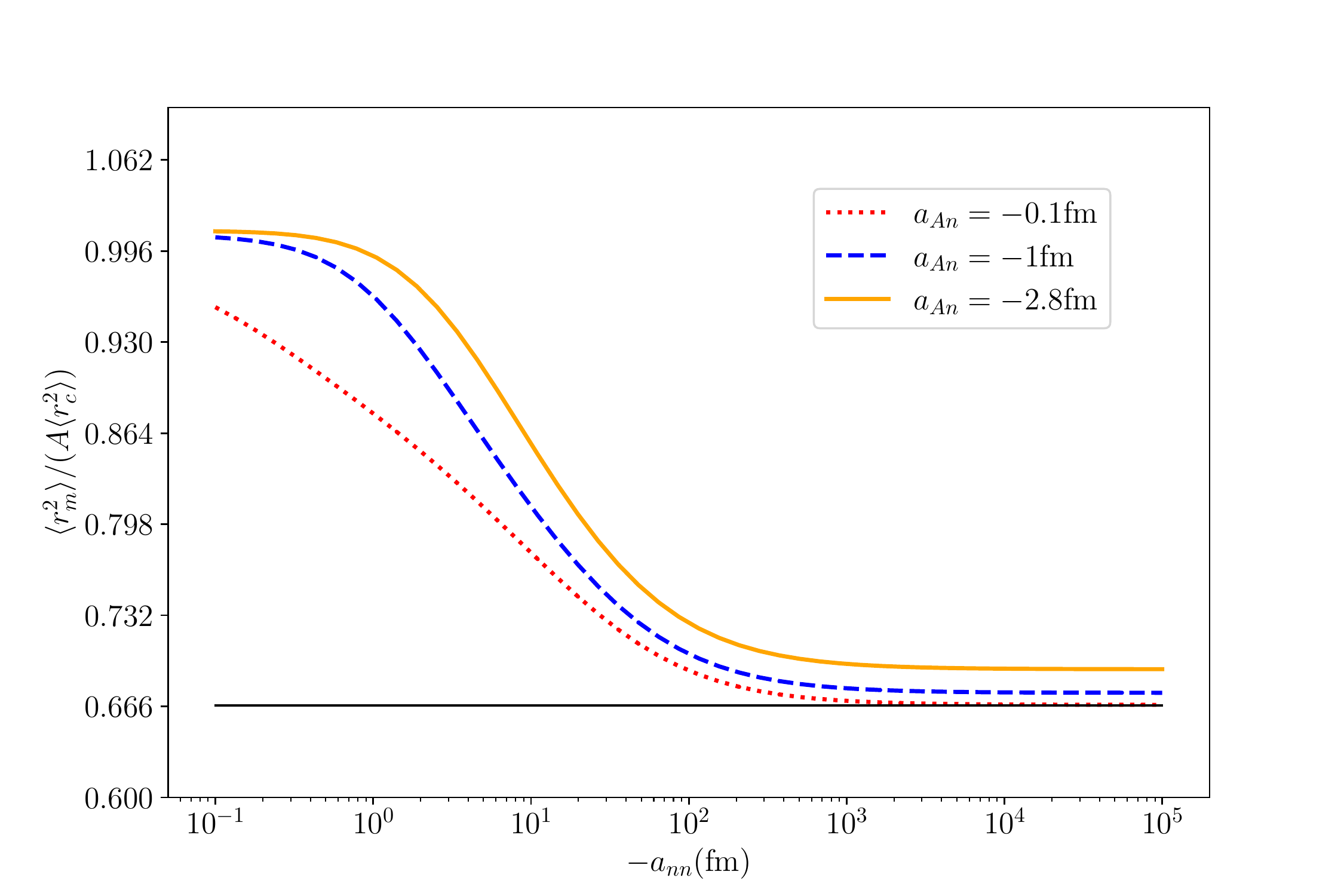}
\par\end{centering}
\caption{\label{fig:unitary-limit}
The ratio $\langle r_m^2\rangle/(A\langle r_c^2\rangle)$ is shown for a few fixed $a_{An}$ values with varying $a_{nn}$. 
The limit $2/3$~\cite{Hongo:2022sdr}, denoted by the black solid reference line, is obtained as $a_{nn}\to-\infty$ and $a_{An}\to 0$. 
Here we set $A=20$ and $B=20$~keV.  }
\end{figure}

The charge, neutron and matter root-mean-square (rms) radii computed using $A=20$, $B=20$~keV and $a_{nn}=-18.6$~fm are shown in Table~\ref{tab:radius}.
On one hand, it becomes clear that the $An$ channel contributes to the radius more significantly as the magnitude of $a_{An}$ increases. 
On the other hand, the universal result in Ref.~\cite{Hongo:2022sdr}, $\langle r_m^2\rangle/(A\langle r_c^2\rangle)$ tending to $2/3$ when $a_{nn}\to-\infty$, is approached in the unitary limit of $a_{nn}\to-\infty$ and at the same time $a_{An}\to 0$ ($a_{An}$ and $a_{nn}$ have the same sign), as shown in Fig. \ref{fig:unitary-limit}.
In that case, the dynamics of the $An$ channel is completely short-ranged and the long-distance physics of the three-body system is dominated by the $nn$ channel. 

\subsection{The results for $^6$He, $^{11}$Li and $^{22}$C }

In the following, we compare the results with Ref.~\cite{Hongo:2022sdr} and pertinent experimental information for two-neutron halo nuclei $^6$He, $^{11}$Li and $^{22}$C. 

For $^6$He, using the binding energy $B=974$~keV, we can reproduce  the universal result 
$\langle r_m^2\rangle/\langle r_c^2\rangle\simeq 0.686A$ in Ref.~\cite{Hongo:2022sdr} in the limit $a_{An}\to 0$. 
The rms charge radius of $^6$He has been precisely measured to be  $(2.054 \pm 0.014)~\mathrm{fm}$~\cite{Wang:2004ze}.
Taking the rms matter radius $(2.59\pm0.05)$~fm extracted in Ref.~\cite{Danilin:2005fh}, one has $\langle r_m^2\rangle/\langle r_c^2\rangle=1.59\pm0.07=(0.40\pm0.02)A$.\footnote{The value of $\langle r_m^2\rangle/\langle r_c^2\rangle$ quoted in Ref.~\cite{Hongo:2022sdr} is $0.862A$ for $^6$He. 
Such a value can be reproduced by adjusting $a_{nn}/a_{An}=3.05$.} 
It is smaller than the unitary limit $2A/3$, and cannot be reproduced at leading order of the NREFT, which leads to values $\geq 2A/3$ as shown in Fig.~\ref{fig:unitary-limit}. 
This is because at leading order for the two-body $An$ interactions defined in Eq.~\eqref{eq:L2}, only a bound or virtual state pole is possible in the $T$ matrix of the $^4$He$\,n$ scattering. 
However, $^5$He is a resonance with a mass 735~keV higher than the $^4$He$\,n$ threshold~\cite{Kondev:2021lzi} and can only be properly treated beyond leading order.
Charge and neutron form factors of two-neutron halo nuclei ${ }^{11} \mathrm{Li}$, ${ }^{14} \mathrm{Be}$ and ${ }^{22} \mathrm{C}$ have been studied at next-to-leading order in halo EFT in Ref.~\cite{Vanasse:2016hgn}.

For $^{11}$Li, the binding energy $B=247$~keV is considered~\cite{Canham:2008jd}. 
In the unitary limit of $a_{nn}\to -\infty$ and setting $a_{An}=0$, the results of $\sqrt{\langle r_c^2\rangle}=0.86$~fm and $\sqrt{\langle r_n^2\rangle}=4.7$~fm in Ref.~\cite{Hongo:2022sdr} can be reproduced by setting the coupling $G_{nn}=0.653$. 
In Ref.~\cite{Hongo:2022sdr}, the coupling is given by a naturalness argument by setting the logarithm in the expression 
\begin{equation}
    g^2(E)=\frac{\pi}{4}\left(\frac{A+2}{A}\right)^{3 / 2} \frac{1}{\ln \frac{E_0}{E}}
\end{equation}
to $1$, where $E_0$ is the energy of the Landau pole of the running coupling, which gives $G_{nn} = g/\sqrt{2}=0.728$. 
However, since this coupling is determined from the wave function in our formalism, we can do the quantitative calculation self-consistently. 
The ground state $^{10}$Li could be either a resonance~\cite{Tilley:2004zz} or a virtual state below the $^9$Li$\,n$ threshold with a virtual energy less than 50~keV~\cite{Garrido:2001ky}. 
We take the latter point of view and set the virtual energy to be 25~keV, and find $a_{^9\text{Li}\,n}=30.3$~fm. 
Correspondingly, we find the rms charge and neutron radii to be $\sqrt{\langle r_c^2\rangle}=0.8$~fm and $\sqrt{\langle r_n^2\rangle}=5.0$~fm, consistent with the values in Ref.~\cite{Canham:2008jd}.

\begin{figure}
\begin{centering}
\includegraphics[scale=0.7]{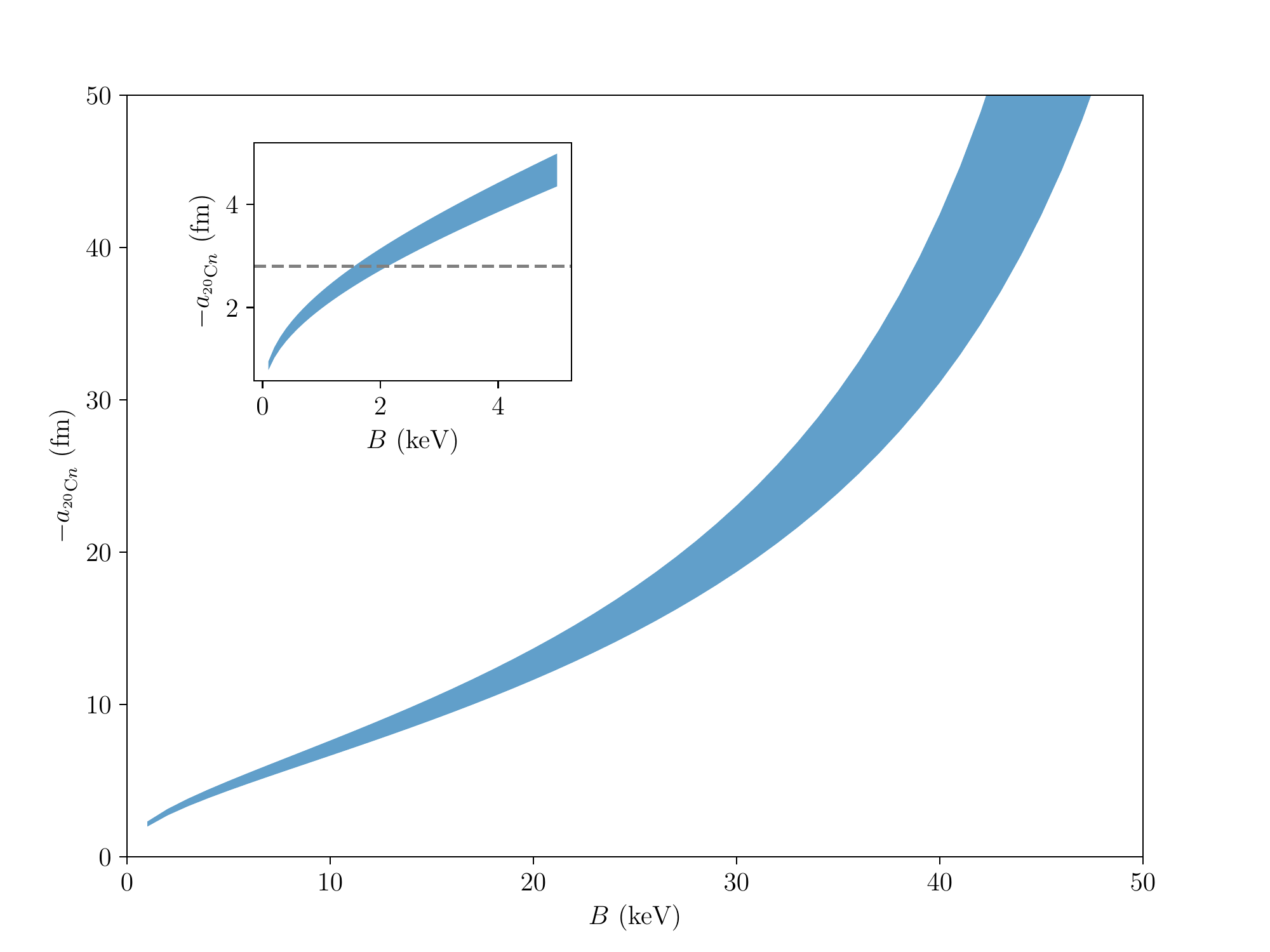}
\par\end{centering}
\caption{\label{fig:allowed-region}
The allowed region for the binding energy (two-neutron separation energy) $B$ of the $^{22}$C and the $S$-wave scattering length between $^{20}\rm{C}$ and $n$, as constrained by the matter radius of $^{22}\rm{C}$, i.e., $r_m=(3.44\pm0.08)$~fm as measured in Ref.~\cite{Togano:2016wyx}. 
The region below the dashed line in the inset is consistent with the constraint $a_{^{20}{\rm C}n}<2.8$~fm set in Ref.~~\cite{Mosby:2013bix}. 
}
\end{figure}

For $^{22}$C,  the rms matter radius was deduced in the experiments as $\sqrt{\langle r_m^2\rangle}=(5.4\pm0.9)$~fm~\cite{Tanaka:2010zza} and $(3.44\pm0.08)$~fm~\cite{Togano:2016wyx}. 
The binding energy of $^{22}$C has a considerable error---the two-neutron separation energy\footnote{The three-body binding energy $B$ of $^{22}$C treated as a two-neutron halo nucleus is equivalent to the two-neutron separation energy $S_{2n}$.} is 
$S_{2n}=(-0.03\pm0.33)$~MeV evaluated from the measured mass excesses of the neutron, $^{20}$C and $^{22}$C, which are 8.07~MeV, $(37.50\pm0.23)$~MeV and $(53.61\pm0.23)$~MeV, respectively~\cite{Kondev:2021lzi}.
Since $^{22}$C is indeed bound, we quote the binding energy as $B=0^{+300}_{-0}$~keV.
The rms matter radius of $^{22}$C measu
red in Ref.~\cite{Tanaka:2010zza} has been used to put constraints on the two-neutron separation energy and the $^{21}{\rm C}$ (as a $^{21}{\rm C}\,n$ system) virtual energy  in Ref.~\cite{Acharya:2013aea}, which suggests $B<100$~keV.
Here we set the binding energy to be $B=20$~keV.
Without the $An$ channel, i.e. when $a_{An}=0$, the rms matter radius is only $1.79$~fm in the unitary limit of $a_{nn}=-\infty$.
The $^{20}\text{C}\,n$ channel is necessary to reproduce the measured values.
We can reproduce $\sqrt{\langle r_m^2\rangle}=5.4$~fm by setting $a_{^{20}\text{C}\,n}=12.5\,a_{nn}=233$~fm and reproduce $\sqrt{\langle r_m^2\rangle}=3.44$~fm by setting $a_{^{20}\text{C}\,n}=0.68a_{nn}=12.6$~fm. 
In both cases, there is a ${^{20}\text{C}\,n}$ virtual state. 
In Fig.~\ref{fig:allowed-region}, we show the allowed region in the $B$-$a_{^{20}{\rm C}\,n}$ plane using the more precise matter radius measurement $(3.44\pm0.08)$~fm~\cite{Togano:2016wyx} as input. 
In Ref.~\cite{Mosby:2013bix}, the absolute value of the $^{20}{\rm C}\,n$ $S$-wave scattering length was constrained to be $<2.8$~fm.
The results imply that the binding energy (two-neutron separation energy) of the $^{22}$C needs to be $\lesssim2$~keV  in order to be simultaneously consistent with the experimental constraints on the matter radius and the $^{20}{\rm C}\,n$ scattering length.
In Table~\ref{tab:radius}, we shows the  dependence of radii on the ratio of couplings, $a_{nn}/a_{An}$. 

The above cases show that the unitary limit results in Ref.~\cite{Hongo:2022sdr} for Borromean nuclei can be reproduced when $|a_{An}|\ll |a_{nn}|$ and by corrections due to the $An$ channel can be computed. 
It is interesting to find that the binding energy, the ratio between the matter and charge radii ($\langle r_m^2\rangle/\langle r_c^2\rangle$), and the ratio between two scattering lengths ($a_{nn}/a_{An}$) are perfectly connected in our formalism.
Thus, the value of $a_{An}$ may be extracted from experimental measurements of the binding energy and $\langle r_m^2\rangle/\langle r_c^2\rangle$ of a two-neutron halo Borromean nucleus.  

\section{\label{sec:Conclusion}Conclusion}

Two-neutron halo nuclei of the Borromean type are studied using halo EFT at leading order.
Both the neutron-neutron dimer and
neutron-core dimer are both introduced in the formalism. 
When the limit of an infinite neutron-neutron scattering length is taken and the neutron-core scattering length is set to vanish, the universal result in Ref.~\cite{Hongo:2022sdr} is
reproduced. 
Corrections from the neutron-core scattering channel to the charge and matter radii of two-neutron halo nuclei are worked out.
Actually, in several realistic examples, such corrections are sizeable.

We find that the shallow three-body bound states like Borromean nuclei can always be accommodated in an Efimov-like interpretation. 
The coupling of the three-body bound state to each of the particle-dimer channels is calculated for various halo nuclei parameters. 
Various physical observables are connected within the formalism.
For example, the three-body binding energy, the ratio between the matter and charge radii ($\langle r_m^2\rangle/\langle r_c^2\rangle$), and the ratio between two scattering lengths ($a_{nn}/a_{An}$) are related.
In this way, the neutron-core scattering length $a_{An}$ may be extracted from various measurements.
We find that the two-neutron separation energy for $^{22}$C needs to be $\lesssim2$~keV in order to be consistent with the experimental constraints of the matter radius of  $^{22}$C~\cite{Togano:2016wyx} and the $^{20}{\rm C}\,n$ $S$-wave scattering length~\cite{Mosby:2013bix}.

\begin{acknowledgments}
The authors thank Chen Ji, Qun Wang and Shan-Gui Zhou for useful discussions. 
This work is supported in part by
the Chinese Academy of Sciences (CAS) under Grants
No. XDB34030000; 
by the National Natural Science Foundation of China (NSFC) under Grant No.~12135011, No.~12125507, No.~11835015, No.~12047503, No.~12175239 and No.~12221005; 
by the NSFC and the Deutsche Forschungsgemeinschaft
(DFG, German Research Foundation) through funds provided to the Sino-German Collaborative Research Center ``Symmetries and the Emergence of Structure in QCD'' (NSFC Grant No. 12070131001, DFG Project-ID 196253076 - TRR110); by the Fundamental Research Funds for the Central Universities; 
and by the National Key R\&D Program of China under Contract No.~2020YFA0406400.

\end{acknowledgments}

\appendix

\section{Normalization of wave function\label{sec:Normalization-of-wave function}}

Let us begin with the scattering equation, 
\begin{align}
\mathcal{M} & =Z+Z\tau\mathcal{M}\,.
\end{align}
In the vicinity of the bound state pole, the amplitude $\mathcal{M}$
has the form in Eq.~\eqref{eq:particle-dimer-Amp}, i.e.,
\begin{align}
\mathcal{M}(\bm p,\bm q;E) & =-\frac{\psi(\bm p)\psi^{\dagger}(\bm q)}{E+B}+\text{regular terms}\,.
\end{align}
Substituting the above decomposition into the identity, 
\begin{align}
\mathcal{M}\mathcal{M}^{-1}\mathcal{M} & =\mathcal{M}\,,
\end{align}
one can find around the pole,
\begin{align}
-\lim_{E\to-B}\psi^{\dagger}\frac{\mathcal{M}^{-1}(E)}{E+B}\psi & =1\,,\label{eq:normalization-derivation}
\end{align}\
where the momentum arguments $\bm p, \bm q$ have been omitted for simplicity.
Since $E=-B$ is the pole energy of the amplitude $\mathcal{M}$, we have
\begin{align}
\mathcal{M}^{-1}(-B) & =0\,.
\end{align}
Using the L'H\^opital rule, one has
\begin{align}
-\lim_{E\to-B}\frac{\mathcal{M}^{-1}(E)}{E+B} =-\frac{\partial}{\partial E}\Big|_{E=-B}\mathcal{M}^{-1}(E)\,.
\end{align}
The inverse amplitude obeys the scattering equation, 
\begin{align}
\mathcal{M}^{-1} & =Z^{-1}-\tau\,,
\end{align}
therefore, 
\begin{align}
-\frac{\partial}{\partial E}\Big|_{E=-B}\mathcal{M}^{-1}(E) & =-\frac{\partial}{\partial E}\Big|_{E=-B}\Big[Z^{-1}-\tau\Big]\nonumber \\
 & =Z^{-1}(-B)\Big(\frac{\partial}{\partial E}\Big|_{E=-B}Z\Big)Z^{-1}(-B)+\Big(\frac{\partial}{\partial E}\Big|_{E=-B}\tau\Big)\nonumber \\
 & =\Big[\tau\Big(\frac{\partial}{\partial E}Z\Big)\tau+\Big(\frac{\partial}{\partial E}\tau\Big)\Big]_{E=-B}\,.
\end{align}
Plugging the result back to Eq.~(\ref{eq:normalization-derivation}),
we will find the normalization, 
\begin{align}
\int\frac{d^{3}\bm p}{(2\pi)^{3}}\frac{d^{3}\bm q}{(2\pi)^{3}}\psi^{\dagger}(\bm p)\left[\tau(\bm p;E)\left(\frac{\partial}{\partial E}Z(\bm p,\bm q;E)\right)\tau(\bm q;E)\right]\Bigg|_{E=-B}\psi(\bm q)\nonumber \\
+\int\frac{d^{3}\bm p}{(2\pi)^{3}}\psi^{\dagger}(\bm p)\left(\frac{\partial}{\partial E}\Bigg|_{E=-B}\tau(\bm p;E)\right)\psi(\bm p)=1\,.
\end{align}

\section{Calculations of Feynman diagrams of the form factors} \label{app:ff}
\subsection{Charge form factor in the $nn$ channel}

\begin{figure}[H]
\begin{centering}
\includegraphics[scale=0.3]{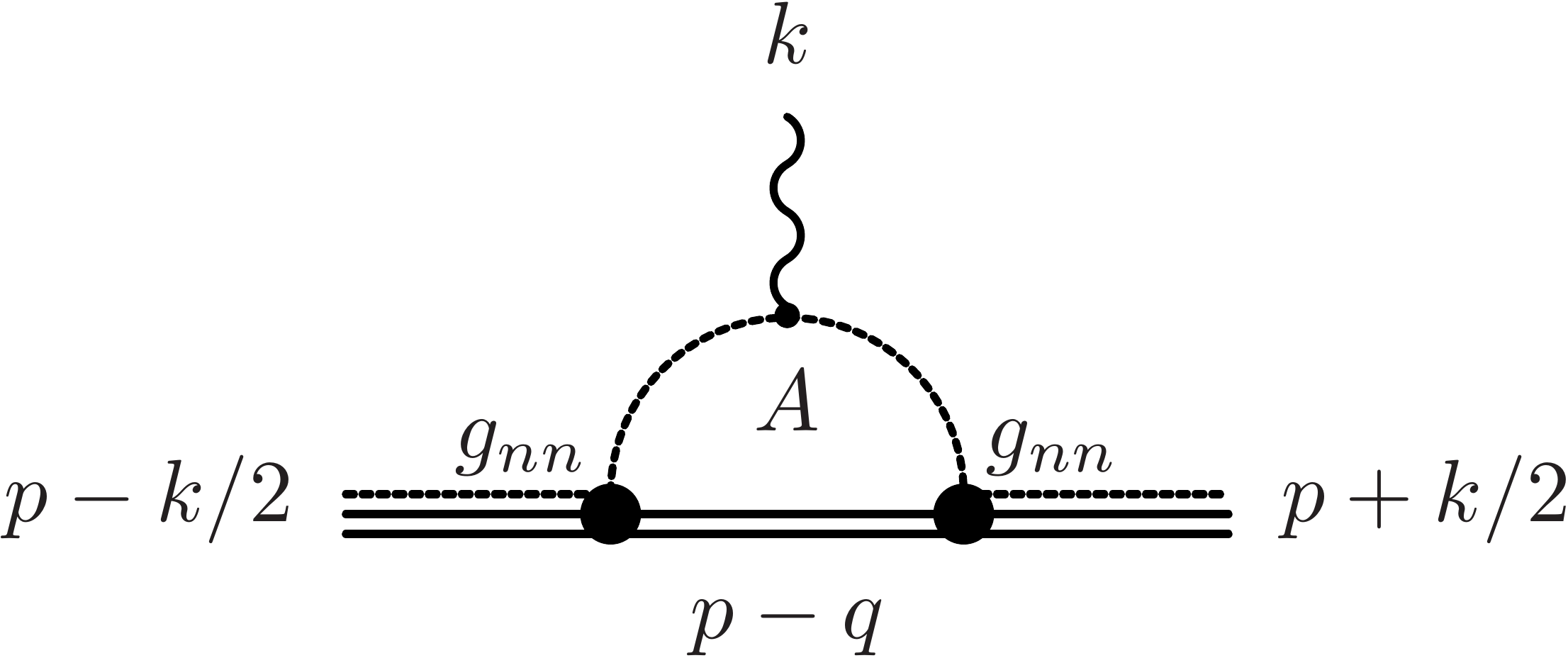}
\par\end{centering}
\caption{Contribution of the $nn$ channel contribution to the charge form factor. \label{fig:charge-nn-channel}}
\end{figure}

We follow the procedure in Ref.~\cite{Zhukov:1993aw}.
The $nn$-channel contribution
to the charge form factor as shown in Fig. \ref{fig:charge-nn-channel}, and it is given by, 
\begin{align}
F_{nn}(k) & =(ig_{nn})^2\int\frac{d^{4}q}{(2\pi)^{4}}\frac{i}{q^{0}-\omega_{A}(\bm{q}-\frac{\bm{k}}{2})+i\epsilon}\frac{i}{q^{0}-\omega_{A}(\bm{q}+\frac{\bm{k}}{2})+i\epsilon}D_{nn}(p-q).
\end{align}
The external momenta are 
\begin{align}
k=(0,\bm{k}),\quad & p=\left(-B+\frac{|\bm{k}|^{2}}{8M},0\right).\label{eq:external-momenta}
\end{align}
Here $M$ is the mass of three-body bound state. 
In the integral, $\omega_{A}$
is the energy of the core $A$, and the propagator of the $nn$ dimer reads,
\begin{align}
D_{nn}(p) & =\frac{8\pi}{im}\,\frac{1}{-a_{nn}^{-1}+\sqrt{p^{2}/4-mp^{0}}}\,.
\end{align}
The result can be expanded as
\begin{align}
F_{nn}(k) & =\mathcal{A}_{nn}+\mathcal{B}_{nn}|\bm{k}|^{2}+O(|\bm{k}|^{4})\,,
\end{align}
where 
\begin{align}
\mathcal{A}_{nn} & =-8\pi g_{nn}^{2}\int\frac{d^{3}q}{(2\pi)^{3}}t_{nn}^{'},\\
\mathcal{B}_{nn} & =-8\pi g_{nn}^{2}\int\frac{d^{3}q}{(2\pi)^{3}}\left[\left(\frac{m}{m_{A}}-\frac{m}{M}\right)\frac{1}{8}t_{nn}^{''}+\frac{m^{2}}{8m_{A}^{2}}\frac{1}{9}\,q^{2}t_{nn}^{'''}\right],
\end{align}
where the core mass is denoted by $m_{A}$ and the functions $t_{nn}$, $t_{nn}^{'}$, $t_{nn}^{''}$ and $t_{nn}^{'''}$ are defined as
\begin{align}
t_{nn} =&\,\frac{1}{-a_{nn}^{-1}+\sqrt{c_{nn}^{2}q^{2}+mB}},\\
t_{nn}^{'} =&\,-\frac{1}{2\sqrt{c_{nn}^{2}q^{2}+mB}}\Bigg(\frac{1}{-a_{nn}^{-1}+\sqrt{c_{nn}^{2}q^{2}+mB}}\Bigg)^{2},\\
t_{nn}^{''} =&\,\frac{1}{2(c_{nn}^{2}q^{2}+mB)}\Bigg(\frac{1}{-a_{nn}^{-1}+\sqrt{c_{nn}^{2}q^{2}+mB}}\Bigg)^{3}\nonumber \\
 & +\frac{1}{4(c_{nn}^{2}q^{2}+mB)^{3/2}}\Bigg(\frac{1}{-a_{nn}^{-1}+\sqrt{c_{nn}^{2}q^{2}+mB}}\Bigg)^{2},\\
t_{nn}^{'''} =&\,-\frac{3}{4(c_{nn}^{2}q^{2}+mB)^{3/2}}\Bigg(\frac{1}{-a_{nn}^{-1}+\sqrt{c_{nn}^{2}q^{2}+mB}}\Bigg)^{4}\nonumber \\
 & -\frac{3}{4(c_{nn}^{2}q^{2}+mB)^{2}}\Bigg(\frac{1}{-a_{nn}^{-1}+\sqrt{c_{nn}^{2}q^{2}+mB}}\Bigg)^{3}\nonumber \\
 & -\frac{3}{8(c_{nn}^{2}q^{2}+mB)^{5/2}}\Bigg(\frac{1}{-a_{nn}^{-1}+\sqrt{c_{nn}^{2}q^{2}+mB}}\Bigg)^{2}.
\end{align}

\subsection{Charge form factor in the $An$ channel}

\begin{figure}[H]
\begin{centering}
\includegraphics[scale=0.3]{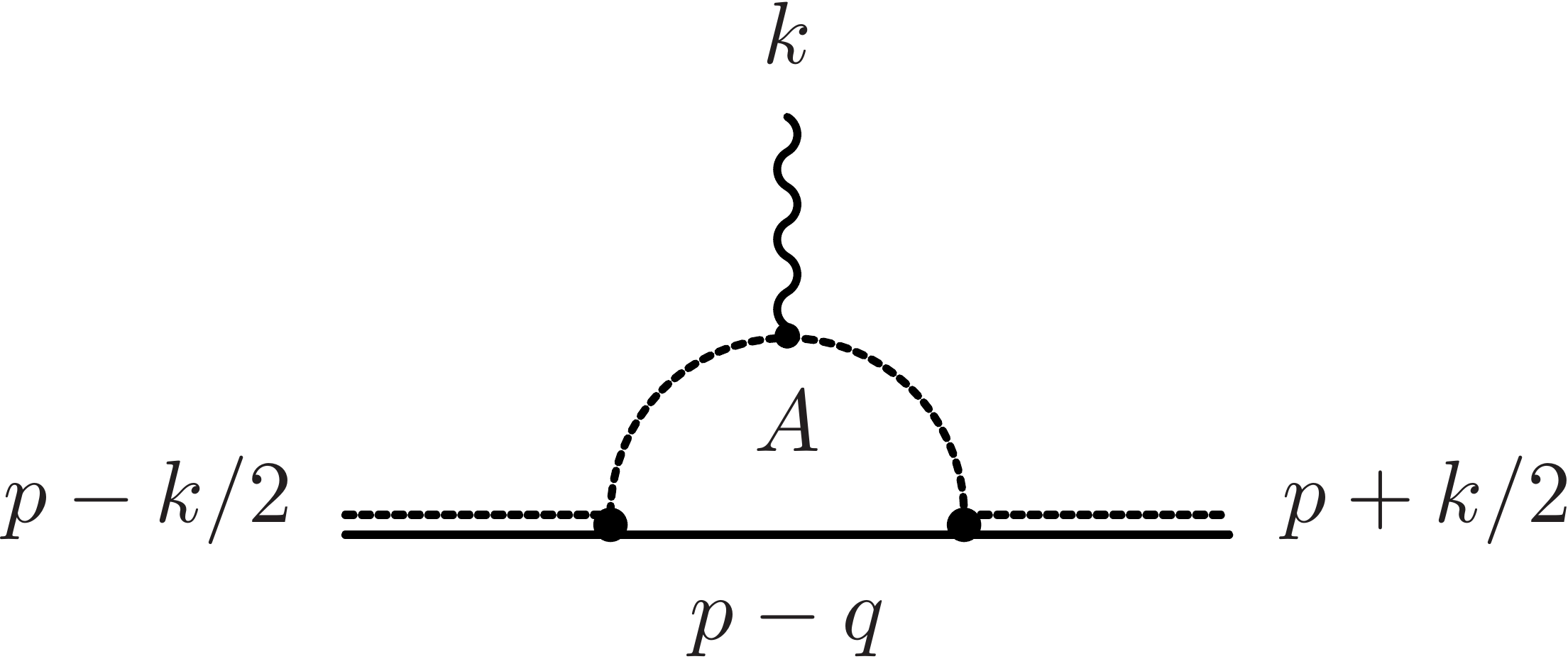}
\par\end{centering}
\caption{Electromagnetic coupling of the $An$ dimer. \label{fig:EM-coupling-An}}
\end{figure}

In order to calculate the contribution from the $An$ channel, we need to resolve the electromagnetic coupling of the $An$ dimer (Fig. \ref{fig:EM-coupling-An}),
denoted by $\Gamma_{An}$, as follows,
\begin{align}
\Gamma_{An}(p,k) & =i^{2}\int\frac{d^{4}q}{(2\pi)^{4}}\frac{i}{q^{0}-\omega_{A}(\bm{q}-\frac{\bm{k}}{2})+i\epsilon}\frac{i}{q^{0}-\omega_{A}(\bm{q}+\frac{\bm{k}}{2})+i\epsilon}\frac{i}{p^{0}-q^{0}-\omega_{n}(\bm{p}-\bm{q})+i\epsilon}
\end{align}
The four momentum $k$ is $(0,\bm{k})$. 
In the integral, we define the energy of neutron $n$ as $\omega_{n}$. 
We adopt the dimensional regularization with the minimal subtraction for the two-body sector, and the result reads 
\begin{align}
\Gamma_{An}(p,k) & =\Gamma_{An}^{(0)}(p)+\Gamma_{An}^{(2)}(p)|\bm{k}|^{2}+\Gamma_{An}^{(2)'}(p)(\bm{p}\cdot\bm{k})^{2}+O(|\bm{k}|^{4})\,,
\end{align}
where 
\begin{align}
\Gamma_{An}^{(0)}(p) & =\frac{\mu^{2}}{2\pi}\frac{1}{\sqrt{c_{f}^{2}p^{2}-2\mu p^{0}}},\\
\Gamma_{An}^{(2)}(p) & =-\frac{\mu^{2}}{4\pi}\frac{d_{f}^{2}}{(c_{f}^{2}p^{2}-2\mu p^{0})^{3/2}},\\
\Gamma_{An}^{(2)'}(p) & =\frac{\mu^{2}}{16\pi}\frac{c_{f}^{4}}{(c_{f}^{2}p^{2}-2\mu p^{0})^{5/2}},
\end{align}
with the coefficients defined as
\begin{align}
c_{f}^{2}=\frac{\mu}{m+m_{A}},\quad & d_{f}^{2}=\frac{\mu}{4m_{A}}-\frac{\mu^{2}}{12m_{A}^{2}}\,.
\end{align}

\begin{figure}[tbh]
\begin{centering}
\includegraphics[scale=0.3]{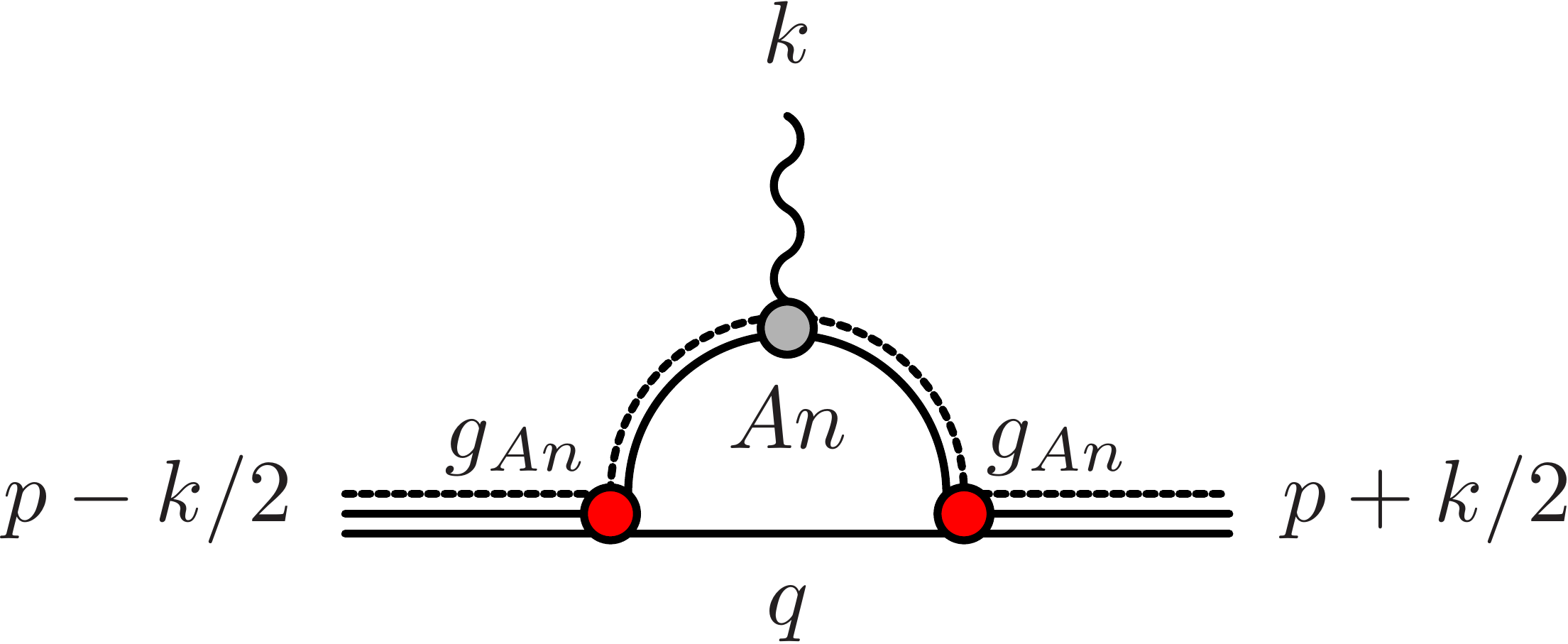}
\par\end{centering}
\caption{Contribution of the $An$ channel to the charge form factor. \label{fig:charge-An-channel}}
\end{figure}

The $An$-channel contribution to the charge form factor of the two-neutron halo nucleus is shown in Fig.~\ref{fig:charge-An-channel}, and it is given by 
\begin{align}
F_{An}(k) & =(ig_{An})^2\int\frac{d^{4}q}{(2\pi)^{4}}D_{An}\left(p-q-\frac{k}{2}\right)\Gamma_{An}(p-q,k)D_{An}\left(p-q+\frac{k}{2}\right)\frac{i}{q^{0}-\omega_{n}(\bm{q})+i\epsilon}.
\end{align}
The external momenta $k$ and $p$ here are the same as those in the $nn$ channel; see Eq.~(\ref{eq:external-momenta}).
In the integral, the propagator of the $An$ dimer reads
\begin{align}
D_{An}(p) & =\frac{2\pi}{i\mu}\,\frac{1}{-a_{An}^{-1}+\sqrt{c_{f}^{2}p^{2}-2\mu p^{0}}}.
\end{align}
The result is 
\begin{align}
F_{An}(k) & =\mathcal{A}_{An}+\mathcal{B}_{An}|\bm{k}|^{2}+O(|\bm{k}|^{4}),
\end{align}
where 
\begin{align}
\mathcal{A}_{An} =&\,2\pi g_{An}^{2}\int\frac{d^{3}q}{(2\pi)^{3}}t_{An}^{2}s_{0}\\
\mathcal{B}_{An} =&\,2\pi g_{An}^{2}\int\frac{d^{3}q}{(2\pi)^{3}}\Bigg\{\frac{1}{2}\left(c_{f}^{2}-\frac{\mu}{M}\right)\left[t_{An}^{'}s_{0}t_{An}\right]+\frac{1}{3}c_{f}^{4}\left[q^{2}t_{An}^{''}s_{0}t_{An}\right]-\frac{1}{3}c_{f}^{4}\left[q^{2}t_{An}^{'2}s_{0}\right]\nonumber \\
 & +t_{An}^{2}s_{0}^{'}+t_{An}^{2}(s_{2}+\frac{1}{3}q^{2}s_{2}^{'})\Bigg\}.
\end{align}
The functions $t_{An}$, $t_{An}^{'}$, $t_{An}^{''}$ and $s_{0}$,
$s_{0}^{'}$, $s_{2}$, $s_{2}^{'}$ are defined as 
\begin{align}
t_{An} =&\,\frac{1}{-a_{An}^{-1}+\sqrt{c_{An}^{2}q^{2}+2\mu B}},\\
t_{An}^{'} =&\,-\frac{1}{2\sqrt{c_{An}^{2}q^{2}+mB}}\Bigg(\frac{1}{-a_{An}^{-1}+\sqrt{c_{An}^{2}q^{2}+2\mu B}}\Bigg)^{2},\\
t_{An}^{''} =&\,\frac{1}{2(c_{An}^{2}q^{2}+2\mu B)}\Bigg(\frac{1}{-a_{An}^{-1}+\sqrt{c_{An}^{2}q^{2}+2\mu B}}\Bigg)^{3}\nonumber \\
 & +\frac{1}{4(c_{An}^{2}q^{2}+2\mu B)^{3/2}}\Bigg(\frac{1}{-a_{An}^{-1}+\sqrt{c_{An}^{2}q^{2}+2\mu B}}\Bigg)^{2},
\end{align}
and
\begin{align}
s_{0} & =\frac{1}{\sqrt{c_{An}^{2}q^{2}+2\mu B}},\\
s_{0}^{'} & =\frac{1}{2}\frac{\mu/(4M)}{(c_{An}^{2}q^{2}+2\mu B)^{3/2}},\\
s_{2} & =-\frac{1}{2}\frac{d_{f}^{2}}{(c_{An}^{2}q^{2}+2\mu B)^{3/2}},\\
s_{2}^{'} & =\frac{1}{8}\frac{c_{f}^{4}}{(c_{An}^{2}q^{2}+2\mu B)^{5/2}}.
\end{align}

Summing up the contributions from both channels and the field strength
coefficient $\sigma_{3}$, we have
\begin{align}
F(k) & =\Big[\sigma_{3}+\mathcal{A}_{nn}+\mathcal{A}_{An}\Big]+\Big[\mathcal{B}_{nn}+\mathcal{B}_{An}\Big]|\bm{k}|^{2}+O(|\bm{k}|^{4}).\label{eq:charge-function}
\end{align}
Due to the field strength normalization, one finds 
\begin{align}
\sigma_{3}+\mathcal{A}_{nn}+\mathcal{A}_{An} & =1.
\end{align}
The second order coefficient in Eq.~(\ref{eq:charge-function}) generates
the charge radius as shown in Eqs.~(\ref{eq:rcnn}) and (\ref{eq:rcAn}) by multiplying the factor $-6$.

\subsection{Neutron form factor in the $nn$ channel}

\begin{figure}[H]
\begin{centering}
\includegraphics[scale=0.3]{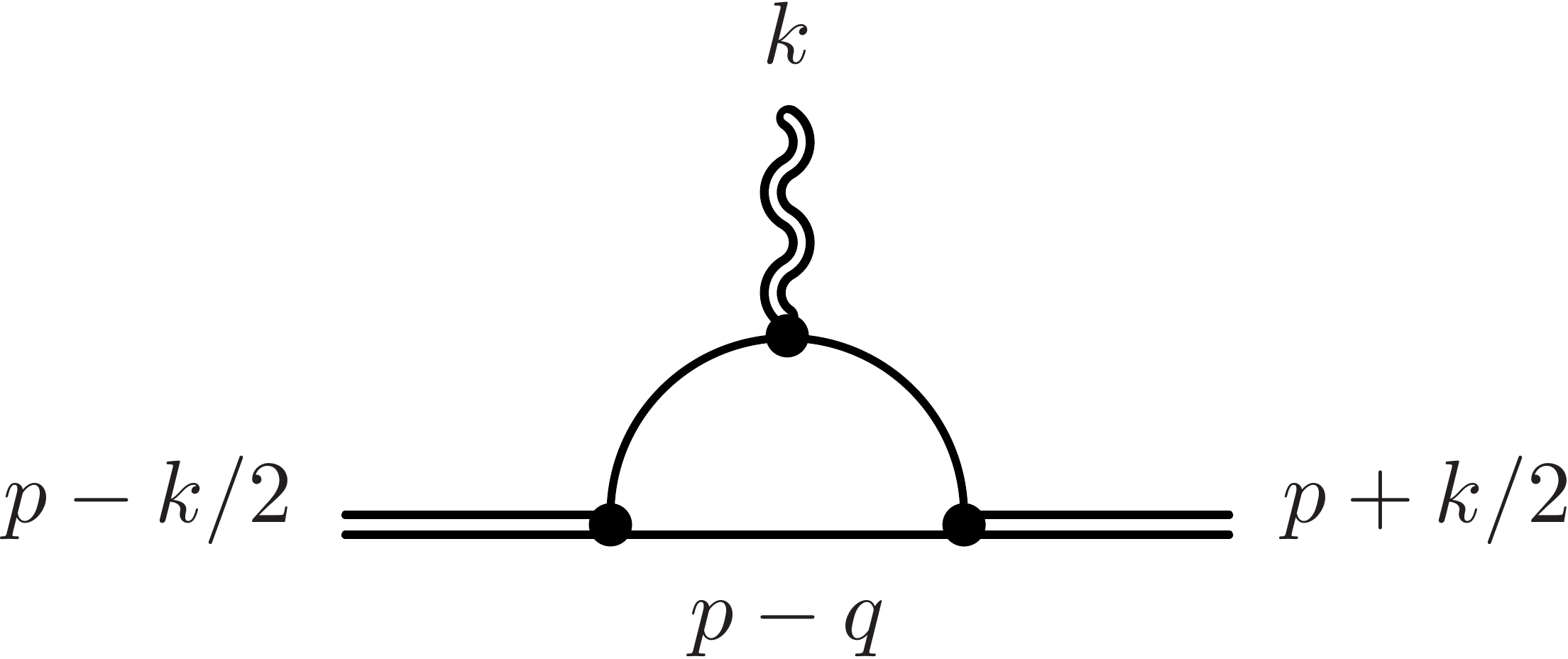}
\par\end{centering}
\caption{Coupling of the neutron number operator to the $nn$ dimer. \label{fig:neutron-coupling-nn}}
\end{figure}

The neutron form factor is calculated in the same way. 
In order to calculate the $nn$-channel contribution as shown in Fig.~\ref{fig:neutron-coupling-nn}, we need to find the coupling of the neutron number operator to the $nn$ dimer, denoted by $\tilde{\Gamma}_{nn}$,  
\begin{align}
\tilde{\Gamma}_{nn}(p,k) & =i^{2}\int\frac{d^{4}q}{(2\pi)^{4}}\frac{i}{q^{0}-\omega_{n}(\bm{q}-\frac{\bm{k}}{2})+i\epsilon}\frac{i}{q^{0}-\omega_{n}(\bm{q}+\frac{\bm{k}}{2})+i\epsilon}\frac{i}{p^{0}-q^{0}-\omega_{n}(\bm{p}-\bm{q})+i\epsilon}.
\end{align}
Analogously, the four momentum $k$ is $(0,\bm{k})$. 
By using the dimensional regularization with minimal subtraction, it is obtained as follows, 
\begin{align}
\tilde{\Gamma}_{nn}(p,k) & =\Gamma_{nn}^{(0)}(p)+\Gamma_{nn}^{(2)}(p)|\bm{k}|^{2}+\Gamma_{nn}^{(2)'}(p)(\bm{p}\cdot\bm{k})^{2}+O(|\bm{k}|^{4}),
\end{align}
where 
\begin{align}
\Gamma_{nn}^{(0)}(p) & =\frac{m^{2}}{8\pi}\frac{1}{\sqrt{p^{2}/4-mp^{0}}},\\
\Gamma_{nn}^{(2)}(p) & =-\frac{5m^{2}}{768\pi}\frac{1}{(p^{2}/4-mp^{0})^{3/2}},\\
\Gamma_{nn}^{(2)'}(p) & =\frac{m^{2}}{1024\pi}\frac{c_{f}^{4}}{(p^{2}/4-mp^{0})^{5/2}}.
\end{align}

\begin{figure}[tbh]
\begin{centering}
\includegraphics[scale=0.3]{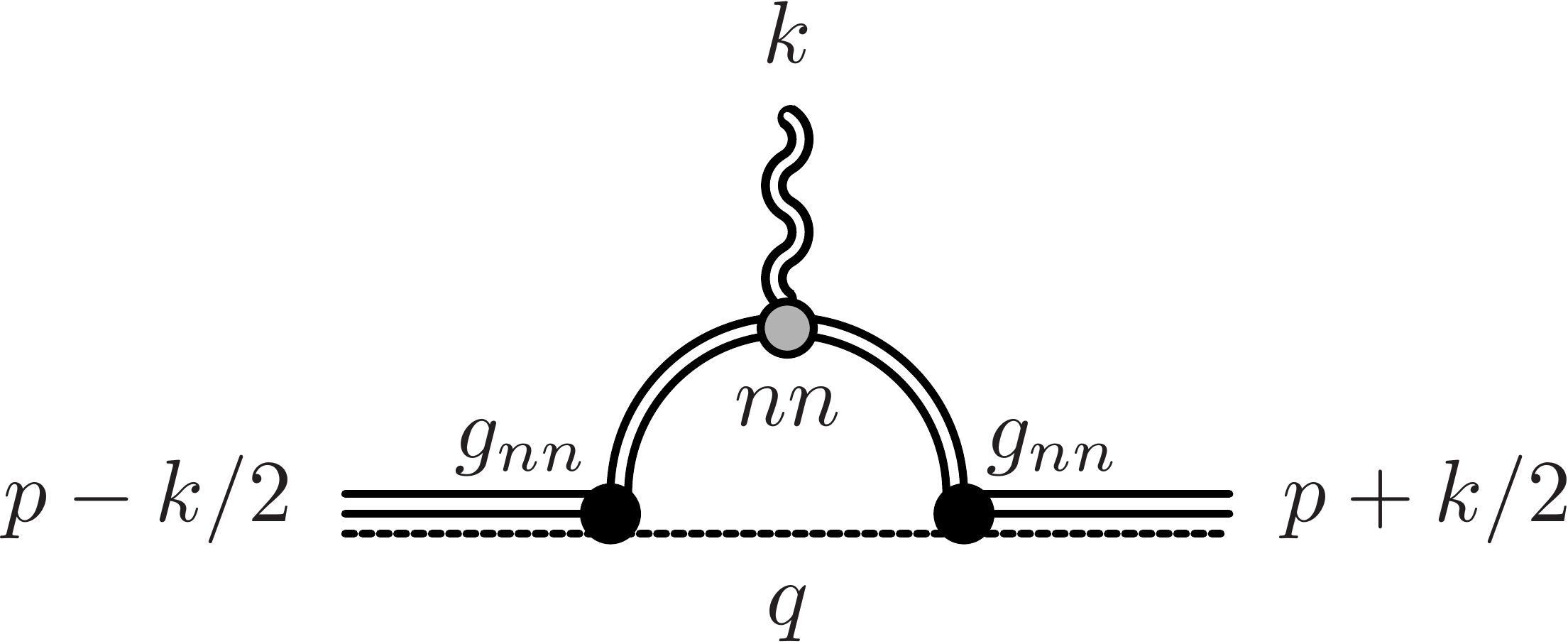}
\par\end{centering}
\caption{Contribution of the $nn$ channel  to the neutron form factor. \label{fig:neutron-nn-channel}}
\end{figure}

The $nn$-channel contribution to the neutron form factor as shown in Fig.
\ref{fig:charge-An-channel}  is given by
\begin{align}
\tilde{F}_{nn}(k) & =(ig_{nn})^{2}\int\frac{d^{4}q}{(2\pi)^{4}}D_{nn}\left(p-q-\frac{k}{2}\right)\tilde{\Gamma}_{nn}(p-q,k)D_{An}\left(p-q+\frac{k}{2}\right)\frac{i}{q^{0}-\omega_{n}(\bm{q})+i\epsilon}.
\end{align}
The external momenta $k$ and $p$ are set in the same way as charge
form factor. 
The result is expanded as 
\begin{align}
\tilde{F}_{nn}(k) & =\tilde{\mathcal{A}}_{nn}+\tilde{\mathcal{B}}_{nn}|\bm{k}|^{2}+O(|\bm{k}|^{4}),
\end{align}
where 
\begin{align}
\tilde{\mathcal{A}}_{nn} =&\,8\pi g_{nn}^{2}\int\frac{d^{3}q}{(2\pi)^{3}}t_{nn}^{2}s_{0,nn},\\
\tilde{\mathcal{B}}_{nn} =&\,8\pi g_{nn}^{2}\int\frac{d^{3}q}{(2\pi)^{3}}\Bigg\{ t_{nn}^{2}\left(s_{0,nn}^{'}+s_{2,nn}+\frac{1}{3}q^{2}s_{2,nn}^{'}\right)\nonumber \\
 & +\left[\left(\frac{1}{8}-\frac{m}{4M}\right)t_{nn}^{'}+\frac{1}{48}q^{2}t_{nn}^{''}\right]s_{0,nn}t_{nn}-\frac{1}{48}q^{2}t_{nn}^{'2}s_{0,nn}\Bigg\}.
\end{align}
The functions $\tilde{s}_{0}$, $\tilde{s}_{0}^{'}$, $\tilde{s}_{2}$
and $\tilde{s}_{2}^{'}$ are defined as
\begin{align}
s_{0,nn} & =\frac{1}{\sqrt{c_{nn}^{2}q^{2}+mB}},\\
s_{0,nn}^{'} & =\frac{1}{2}\frac{m/(8M)}{(c_{nn}^{2}q^{2}+mB)^{3/2}},\\
s_{2,nn} & =-\frac{5}{96}\frac{1}{(c_{nn}^{2}q^{2}+mB)^{3/2}},\\
s_{2,nn}^{'} & =\frac{1}{128}\frac{1}{(c_{nn}^{2}q^{2}+mB)^{5/2}}.
\end{align}

\subsection{Neutron form factor in the $An$ channel}

\begin{figure}[H]
\begin{centering}
\includegraphics[scale=0.3]{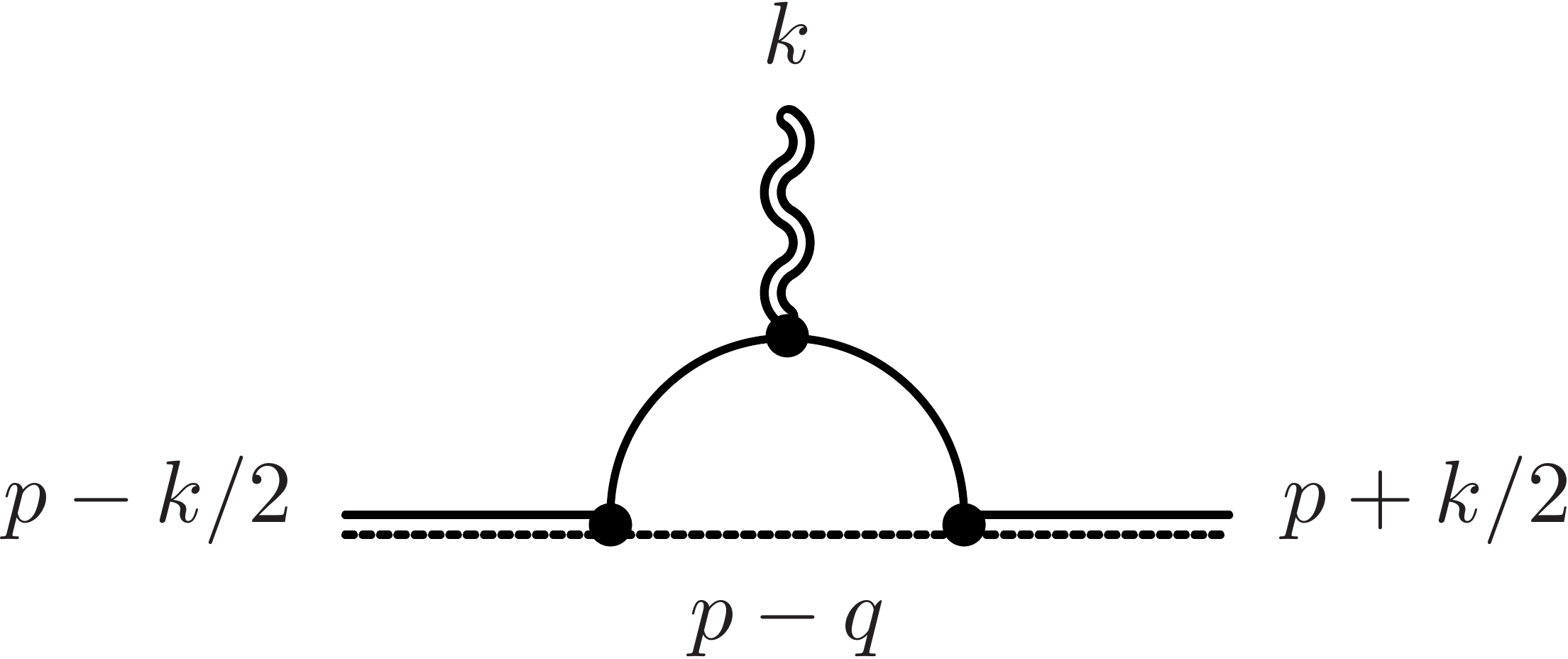}
\par\end{centering}
\caption{Coupling of the neutron number operator to the $An$ dimer. \label{fig:neutron-coupling-An}}
\end{figure}

In order to calculate the $An$-channel contribution as shown in Fig. \ref{fig:neutron-coupling-An}, the neutron
coupling of the $An$ dimer, denoted
by $\tilde{\Gamma}_{An}$, should be calculated firstly. It is given by
\begin{align}
\tilde{\Gamma}_{An}(p,k) & =i^{2}\int\frac{d^{4}q}{(2\pi)^{4}}\frac{i}{q^{0}-\omega_{n}(\bm{q}-\frac{\bm{k}}{2})+i\epsilon}\frac{i}{q^{0}-\omega_{n}(\bm{q}+\frac{\bm{k}}{2})+i\epsilon}\frac{i}{p^{0}-q^{0}-\omega_{A}(\bm{p}-\bm{q})+i\epsilon}.
\end{align}
The four-momentum $k$ is $(0,\bm{k})$. 
The result is 
\begin{align}
\tilde{\Gamma}_{An}(p,k) & =\Gamma_{An}^{(0)}(p)+\tilde{\Gamma}_{An}^{(2)}(p)|\bm{k}|^{2}+\Gamma_{An}^{(2)'}(p)(\bm{p}\cdot\bm{k})^{2}+O(|\bm{k}|^{4}),
\end{align}
where 
\begin{align}
\tilde{\Gamma}_{An}^{(2)}(p) & =-\frac{\mu^{2}}{4\pi}\frac{\tilde{d}_{f}^{2}}{(c_{f}^{2}p^{2}-2\mu p^{0})^{3/2}}.
\end{align}
Here we have defined an additional coefficient as follows,
\begin{align}
\tilde{d}_{f}^{2} & =\frac{\mu}{4m}-\frac{\mu^{2}}{12m^{2}}.
\end{align}

\begin{figure}[H]
\begin{centering}
\includegraphics[width=0.45\textwidth]{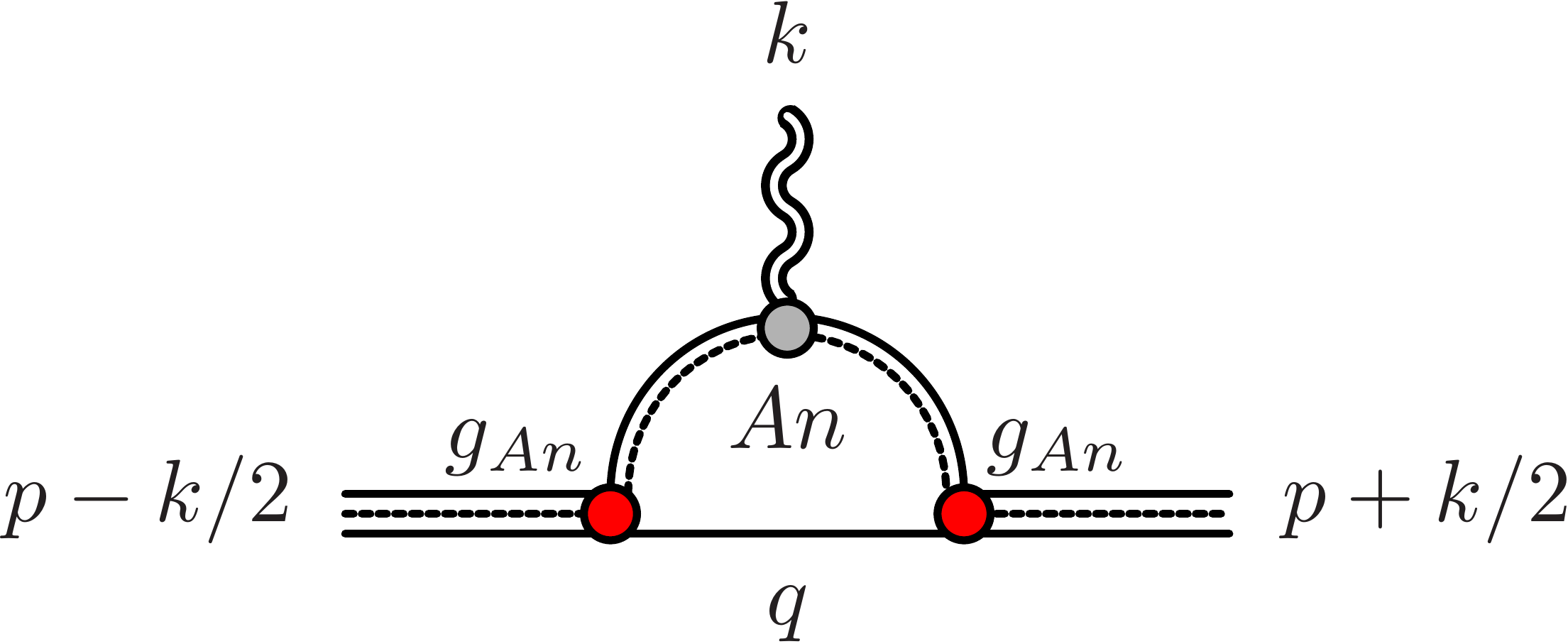}\hfill 
\includegraphics[width=0.45\textwidth]{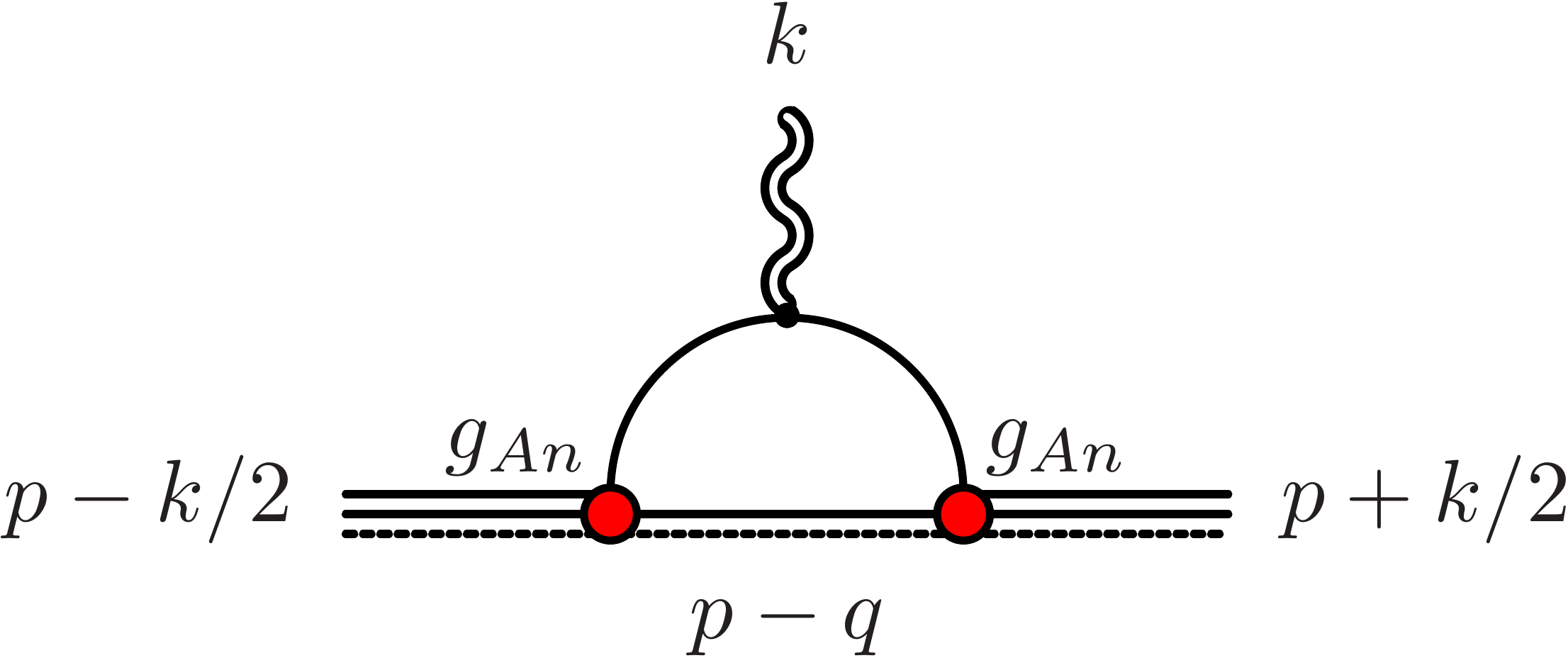}
\end{centering}
\caption{Contribution of the $An$ channel to the neutron form factor. \label{fig:neutron-An-channel}}
\end{figure}

The $An$-channel contribution to the neutron form factor as shown in Fig. \ref{fig:neutron-An-channel} contains
two terms. The first term $\tilde{F}_{An,1}$
is given by
\begin{align}
\tilde{F}_{An,1}(k) & =(ig_{An})^{2}\int\frac{d^{4}q}{(2\pi)^{4}}D_{An}\left(p-q-\frac{k}{2}\right)\tilde{\Gamma}_{An}(p-q,k)D_{An}\left(p-q+\frac{k}{2}\right)\frac{i}{q^{0}-\omega_{n}(\bm{q})+i\epsilon}.
\end{align}
 We have the same momenta setting as in Eq.~(\ref{eq:external-momenta}).
The result is expanded as 
\begin{align}
\tilde{F}_{An,1}(k) & =\tilde{\mathcal{A}}_{An,1}+\tilde{\mathcal{B}}_{An,1}|\bm{k}|^{2}+O(|\bm{k}|^{4}),
\end{align}
where 
\begin{align}
\tilde{\mathcal{A}}_{An,1} =&\,2\pi g_{An}^{2}\int\frac{d^{3}q}{(2\pi)^{3}}t_{An}^{2}s_{0},\\
\tilde{\mathcal{B}}_{An,1} =&\,2\pi g_{An}^{2}\int\frac{d^{3}q}{(2\pi)^{3}}\Bigg\{ t_{An}^{2}\left(s_{0}^{'}+\tilde{s}_{2}+\frac{1}{3}q^{2}s_{2}^{'}\right)+
\nonumber\\
 & +\left[\frac{1}{2}\left(c_{f}^{2}-\frac{\mu}{M}\right)t_{An}t_{An}^{'}+\frac{1}{3}c_{f}^{4}\left(t_{An}t_{An}^{''}-t_{An}^{'2}\right)q^{2}\right]s_{0}\Bigg\}.
\end{align}
with
\begin{align}
\tilde{s}_{2} & =-\frac{1}{2}\frac{\tilde{d}_{f}^{2}}{\left(c_{An}^{2}q^{2}+2\mu B\right)^{3/2}}.
\end{align}

The second term $\tilde{F}_{An,2}$ is given by
\begin{align}
\tilde{F}_{An,2}(k) & =(ig_{An})^{2}\int\frac{d^{4}q}{(2\pi)^{4}}\frac{i}{q^{0}-\omega_{n}(\bm{q}-\frac{\bm{k}}{2})+i\epsilon}\frac{i}{q^{0}-\omega_{n}(\bm{q}+\frac{\bm{k}}{2})+i\epsilon}D_{An}(p-q),
\end{align}
and the result reads 
\begin{align}
\tilde{F}_{An,2}(k) & =\tilde{\mathcal{A}}_{An,2}+\tilde{\mathcal{B}}_{An,2}|\bm{k}|^{2}+O(|\bm{k}|^{4}),
\end{align}
where 
\begin{align}
\tilde{\mathcal{A}}_{An,2} & =-4\pi g_{An}^{2}\int\frac{d^{3}q}{(2\pi)^{3}}t_{An}^{'},\\
\tilde{\mathcal{B}}_{An,2} & =-4\pi g_{An}^{2}\int\frac{d^{3}q}{(2\pi)^{3}}\Bigg[\left(\frac{\mu}{4m}-\frac{\mu}{4M}\right)t_{An}^{''}+\frac{\mu^{2}}{18m^{2}}t_{An}^{'''}q^{2}\Bigg].
\end{align}
Here we define additionally 
\begin{align}
t_{An}^{'''} =&\,-\frac{3}{4(c_{An}^{2}q^{2}+2\mu B)^{3/2}}\Bigg(\frac{1}{-a_{An}^{-1}+\sqrt{c_{An}^{2}q^{2}+2\mu B}}\Bigg)^{4}\nonumber \\
 & -\frac{3}{4(c_{An}^{2}q^{2}+2\mu B)^{2}}\Bigg(\frac{1}{-a_{An}^{-1}+\sqrt{c_{An}^{2}q^{2}+2\mu B}}\Bigg)^{3}\nonumber \\
 & -\frac{3}{8(c_{An}^{2}q^{2}+2\mu B)^{5/2}}\Bigg(\frac{1}{-a_{An}^{-1}+\sqrt{c_{An}^{2}q^{2}+2\mu B}}\Bigg)^{2}.
\end{align}

We put all the contributions to the neutron form factor together and obtain
\begin{align}
F_n(k) =\Big[2\sigma_{3}+\tilde{\mathcal{A}}_{nn}+\tilde{\mathcal{A}}_{An,1}+\tilde{\mathcal{A}}_{An,2}\Big]+\Big[\tilde{\mathcal{B}}_{nn}+\tilde{\mathcal{B}}_{An,1}+\tilde{\mathcal{B}}_{An,2}\Big]|\bm{k}|^{2}+O(|\bm{k}|^{4}).\label{eq:neutron-function}
\end{align}
It can be checked that the free neutron number of a two-neutron halo nucleus is exactly
$2$ after normalization, i.e., 
\begin{align}
2\sigma_{3}+\tilde{\mathcal{A}}_{nn}+\tilde{\mathcal{A}}_{An,1}+\tilde{\mathcal{A}}_{An,2} & =2.
\end{align}
Then we can multiply the second order coefficient in Eq.~(\ref{eq:neutron-function}) by a factor $-6$ and get the neutron radius as shown in Eqs.~(\ref{eq:rnnn}) and (\ref{eq:rnAn}).

\bibliography{borromeanrefs.bib}

\end{document}